\documentclass[acmlarge]{acmart}

\setcopyright{acmlicensed}
\acmJournal{IMWUT}
\acmYear{2022} \acmVolume{6} \acmNumber{1} \acmArticle{16}
\acmMonth{3} \acmPrice{15.00}\acmDOI{10.1145/3517232}

\usepackage{xspace}
\usepackage{booktabs}
\usepackage{url}
\usepackage{color}
\usepackage{caption}
\usepackage{subcaption}
\newcommand{\avP}[0]{Voice Assistants\xspace}
\newcommand{\avS}[0]{Voice Assistant\xspace}

\newcommand{\sys}[0]{\textsf{SkillFence}\xspace}
\newcommand{\appv}[0]{skill\xspace}
\newcommand{\appvs}[0]{skills\xspace}

\newcommand{\mjrev}[1]{{#1}}

\newcommand{\numUsersTotal}[0]{$116$\xspace}

\newcommand{\alexaSameName}[0]{$7656$\xspace}
\newcommand{\alexaSameInvoke}[0]{$8978$\xspace}
\newcommand{\totalSkills}[0]{$51964$\xspace}
\newcommand{\acLinkingSkills}[0]{$3659$\xspace}

\newcommand{\coverage}[0]{$90.83\%$\xspace}
\newcommand{\far}[0]{$19.83\%$\xspace}

\begin{document}

\title{SkillFence: A Systems Approach to Practically Mitigating Voice-Based Confusion Attacks}


\author{Ashish Hooda}
\orcid{0000-0002-2928-919X}
\affiliation{
    \institution{University of Wisconsin--Madison}
    \streetaddress{1210 West Dayton St}
    \city{Madison}
    \state{Wisconsin}
    \country{USA}
}
\email{ahooda@wisc.edu}

\author{Matthew Wallace}
\orcid{0000-0002-9796-7247}
\affiliation{
    \institution{University of Wisconsin--Madison}
    \streetaddress{1415 Engineering Dr}
    \city{Madison}
    \state{Wisconsin}
    \country{USA}
}
\email{mhwallace@wisc.edu}

\author{Kushal Jhunjhunwalla}
\orcid{0000-0001-5841-8459}
\affiliation{
    \institution{University of Washington}
    \streetaddress{185 E Stevens Way NE}
    \city{Seattle}
    \state{Washington}
    \country{USA}
}
\email{kushaljh@cs.washington.edu}

\author{Earlence Fernandes}
\orcid{0000-0001-8593-2840}
\affiliation{
    \institution{University of Wisconsin--Madison}
    \streetaddress{1210 West Dayton St}
    \city{Madison}
    \state{Wisconsin}
    \country{USA}
}
\email{earlence@cs.wisc.edu}

\author{Kassem Fawaz}
\orcid{0000-0002-4609-7691}
\affiliation{
    \institution{University of Wisconsin--Madison}
    \streetaddress{1415 Engineering Dr}
    \city{Madison}
    \state{Wisconsin}
    \country{USA}
}
\email{kfawaz@wisc.edu}

%\author{Lars Th{\o}rv{\"a}ld}
%\affiliation{%
%  \institution{The Th{\o}rv{\"a}ld Group}
%  \streetaddress{1 Th{\o}rv{\"a}ld Circle}
%\email{larst@affiliation.org}

%\affiliation{%
%  \institution{Inria Paris-Rocquencourt}

%\affiliation{%
% \institution{Rajiv Gandhi University}

%\affiliation{%
%  \institution{Tsinghua University}
%  \streetaddress{30 Shuangqing Rd}

%\affiliation{%
%  \institution{Palmer Research Laboratories}
%  \streetaddress{8600 Datapoint Drive}
%  \country{USA}

%\affiliation{%
%  \institution{The Th{\o}rv{\"a}ld Group}
%  \streetaddress{1 Th{\o}rv{\"a}ld Circle}
%\email{jsmith@affiliation.org}

%\affiliation{%
%  \institution{The Kumquat Consortium}
%  \city{New York}
%\email{jpkumquat@consortium.net}

\renewcommand{\shortauthors}{Hooda et al.}
\begin{abstract}

%Voice assistants rely on accurate speech recognition and an open market of voice-based apps (i.e., skills) to provide useful functionality to users, ranging from controlling physical devices to ordering a cab. 

\mjrev{
Voice assistants are deployed widely and provide useful functionality. However, recent work has shown that commercial systems like Amazon Alexa and Google Home are vulnerable to voice-based confusion attacks that exploit design issues. We propose a systems-oriented defense against this class of attacks and demonstrate its functionality for Amazon Alexa. We ensure that only the skills a user intends execute in response to voice commands. Our key insight is that we can interpret a user's intentions by analyzing their activity on counterpart systems of the web and smartphones. For example, the Lyft ride-sharing Alexa skill has an Android app and a website. Our work shows how information from counterpart apps can help reduce dis-ambiguities in the skill invocation process. We build \sys, a browser extension that existing voice assistant users can install to ensure that only legitimate skills run in response to their commands. Using real user data from MTurk ($N=$~\numUsersTotal) and experimental trials involving synthetic and organic speech, we show that \sys\xspace provides a balance between usability and security by securing \coverage of skills that a user will need with a False acceptance rate of \far}.
\end{abstract}


%\begin{CCSXML}
%<ccs2012>
%  <concept_id>10010520.10010553.10010562</concept_id>
%  <concept_desc>Computer systems organization~Embedded systems</concept_desc>
%  <concept_significance>500</concept_significance>
%  <concept_id>10010520.10010575.10010755</concept_id>
%  <concept_desc>Computer systems organization~Redundancy</concept_desc>
%  <concept_significance>300</concept_significance>
%  <concept_id>10010520.10010553.10010554</concept_id>
%  <concept_desc>Computer systems organization~Robotics</concept_desc>
%  <concept_significance>100</concept_significance>
%  <concept_id>10003033.10003083.10003095</concept_id>
%  <concept_desc>Networks~Network reliability</concept_desc>
%  <concept_significance>100</concept_significance>
%</ccs2012>
%\end{CCSXML}

\maketitle

\section{Introduction}
\label{intro}
Rising in popularity, voice assistants, like Amazon Alexa and Google Home, help users accomplish a variety of tasks using speech as input. Through integrating third-party applications, called skills in Alexa terminology, voice assistants empower their users to access personal information, control physical devices in homes, and perform financial transactions. Although these devices are useful, they carry significant security and privacy risks to their users.

% Discuss what these attacks are
\mjrev{Beyond traditional computer security vulnerabilities, recent work has shown that voice assistants are vulnerable to \textit{voice-based confusion attacks}~\cite{kumar,zhang,lipfuzzer}. To improve their usability, \avP auto-enable \appvs to directly execute after the user speaks~\cite{alexa-no-perm-prompt}. Because of imperfections in the speech interpretation pipeline, the \avS might execute a skill that the user did not intend or expect. Consider a \appv that allows users to interact with the Fitbit account. A malicious \appv might have the name \textit{Phitbit}. The speech recognition systems can confuse both skills, and invoke the malicious \appv, instead of the true fitness tracking Fitbit skill, without the user noticing. Once activated, such a \appv can steal private information~\cite{shezan-www2020} or perform other dangerous actions. Similar attacks mimic the voice interface of legitimate \appvs~\cite{zhang} or rely on semantic interpretation errors of natural language understanding~\cite{lipfuzzer}. 
}

% why do they exist?
We identify several fundamental deficiencies in the design of systems like Alexa that contribute to these attacks. First, natural language is ambiguous and speech recognition is prone to errors. Second, there is loose vetting of skill metadata allowing attackers to impersonate legitimate skills. Third, the traditional security and privacy feedback loop with the user is severely hampered because voice is often the only input-output modality~\cite{major2019alexa}.

% Third, voice-only interaction is fundamentally limited compared to standard computers, leading to a break of the typical visual feedback loop --- it is easy to mislead users~\cite{major2019alexa}.

%  We concurrently contribute the same findings by getting an attack skill `PhitBit' \mjrev{certified by} the Alexa marketplace that is designed to squat on FitBit. 
 
% what have people tried to fix the issue? and where do they fall short?
Independently approaching each of these deficiencies leads to incomplete protection against voice confusion attacks. First, while improving the speech interpretation process leads to fewer errors, there are fundamentally ambiguous situations that are impossible to resolve without any additional information (e.g., the difference between Fitbit and Phitbit). Second, Amazon's vetting of the skills is inadequate. While Amazon's guidelines for skill developers prohibits them from using phonetically-close skill invocation phrases, we demonstrate -- similar to recent work~\cite{alexavetting, hu2020case} -- that attackers can violate these policies and get their skills published to the market~\cite{alexavetting, hu2020case}. We are able to certify a ``Phitbit'' skill to squat on the Fitbit skill on the Alexa marketplace\footnote{More discussion about the ethical issues around certifying this attack skill are in Section~\ref{insufficient-vet}}. An attacker can also mirror the metadata of a legitimate skill, such as its invocation phrase, name, privacy policy URL and account linking URL implying that there is no distinguishing information between an attacker and legitimate skill.  Voice assistant vendors could prevent skills with identical or phonetically-similar invocation phrases from being uploaded, but this limits flexibility and openness of the market and can lead to squatting behavior that is common on the web.  For example, someone could register a skill for Fitbit and then ask the fitness tracking company to pay a large sum of money for that invocation phrase. There are also legitimate reasons for why skills have similar invocation phrases, such as multiple skills for the same task (Section~\ref{insufficient-vet}). Finally, involving the user in the execution loop of each skill hinders usability --- as discussed earlier, Amazon Alexa now provides auto-enabling of skills for enhanced usability.

% Why Amazon cannot police their market, brief intro...
%\textcolor{blue}{From a central entity's perspective, policing the Alexa Marketplace is a difficult proposition. Attempting to monitor millions of Alexa skill libraries creates the possibility of large-scale privacy breaches. Monitoring the phonetic space for attempted skill squatting instances creates a scenario that discourages developer interest and quashes Marketplace growth. Section \ref{sec:amazon_policing}'s \textbf{Amazon Policing} provides an in-depth explanation of the problems faced by a central entity when attempting to protect users and encourage developers simultaneously.} 

% what we do about it; what is our key insight.
\mjrev{We propose \sys, which adopts a systems-view of the problem and incorporates the user's preferences when speaking a voice command. Our key insight is that we can disambiguate skills with similar invocation phrases by analyzing the user's activity on \emph{counterpart systems,} especially web and smartphone apps. Taking our example above, \textit{Fitbit} has a website and a mobile app in addition to the Amazon Alexa skill. Most users will have already used these websites and smartphone apps before they turn to use skills. For some of them, the user has to sign up for the service through a website or smartphone app. Therefore, by noticing that a user has visited the Fitbit website consistently or has downloaded the corresponding Android app, we can select \textit{Fitbit} (instead of \textit{Phitbit}) as the correct skill when the user speaks the phrase ``open Fitbit''.}

% After predicting the set of skills a user might need, we enable them while disabling the phonetically-close ones. This has the effect of biasing Alexa to always execute the enabled skill instead of other ones.

% the challenges in our approach and how we overcome
% securely linking counterpart acitivty to amazon skill
% ensuring that once a skill is predicted, only that predicted one is executed
% making it work using commodity alexa: proprietary system, protect users today
\mjrev{Achieving \sys's design requires overcoming several challenges: (1) Once we identify a user's counterpart activity, it must be \emph{securely} matched to the corresponding Alexa skills. As we discussed, an attacker can mirror the metadata of legitimate skills. Thus, skills in the Alexa ecosystem do not have a notion of \emph{secure identity} --- a fundamental design principle in other computing systems (e.g., websites have TLS certificates and Android apps are signed).  (2) Once we match a skill to its counterpart securely, we must ensure that only the matched skill gets invoked in response to an ambiguous phrase and despite the presence of phonetically-similar skills on the market. (3) We must achieve these two properties on the proprietary Alexa ecosystem over which we have no control so that we can protect users today.}

\begin{figure}[t]
    \centering
    \includegraphics[width=0.55\columnwidth]{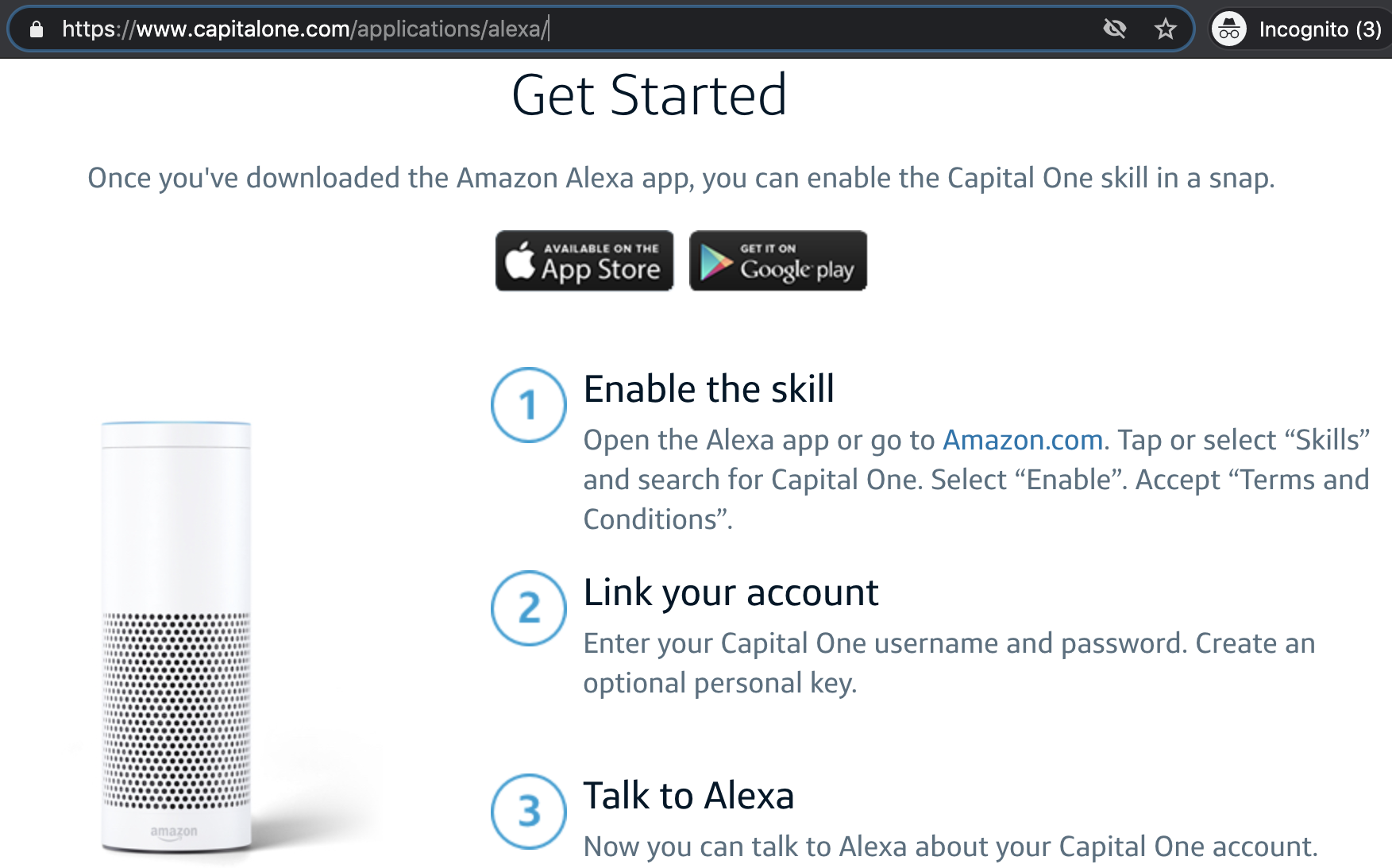}
    \caption{Capital One skill information on its website. We extend the identity of the website to the Alexa skill if a link to it is found on a website we derive from the skill's metadata.}
    \label{fig:cap-one}
\end{figure}

\sys overcomes these challenges using a series of insights. First, we develop a secure identity for Alexa skills \emph{without} modifying the proprietary platform. Our insight is that skill developers tend to link their Alexa skills directly from a website they own, much like they link their iOS and Android apps. For example, Capital One places a link to its Amazon skill on its website (Figure~\ref{fig:cap-one}), like it does for its Android and iOS apps. Assuming that Capital One's website is not compromised, we obtain a secure link to the corresponding skill. Thus, we extend the user's trust in the website to the skill itself. Our system automatically discovers such links (Section~\ref{secure_backlinking}).  

The second challenge involves ensuring that securely \mjrev{matched} skills execute in response to ambiguous commands. We discover that the Alexa backend provides two APIs --- enable a skill and disable a skill --- that can be used to control Alexa's invocation behavior. Specifically, we find that if a skill is enabled and its phonetic-neighbors are disabled, Alexa executes the enabled skill. We characterize and confirm this behavior using large-scale voice experiments (Section~\ref{eval_enable_disable}). 

% need to rewrite the below actually. 
% We study  \sys with a tunable knob for the user to balance the security-usability trade-off.
\mjrev{
We implement \sys as a browser extension and a backend server. We evaluate it by designing and launching a data collection effort. Using the data of \numUsersTotal users, our trace-based evaluation finds that \sys can select the correct skill (the one user intends to use) in \coverage of the cases. We then evaluate \sys in the live Alexa environment by performing real time invocations for each skill used by the users. For those selected skills selected, \sys reduces the incorrect invocations from $17$ to $0$.
% Section 5.4 - using end to end invocations we find that there are no incorrect invocations for skills that SF can identify/match
%Furthermore, through enabling the matched skills and disabling their phonetic neighbors, \sys reduces incorrect skill invocations by $74\%$.
}

\mjrev{
Finally, we provide an evidence-based set of design recommendations distilled from our experience in building and evaluating \sys. If implemented by stakeholders, they will secure the Alexa ecosystem and further improve the efficacy of \sys. For example, a skill vendor can place a link on its website that connects to the Amazon webpage listing of its skill, permitting \sys to automatically associate the identity of the website with the skill. We have started outreach efforts in explaining our recommendations to skill developers and are starting to see adoption --- a skill developer has already updated their website to include a link to the Alexa skill, therefore helping \sys compute a secure identity for it (Section~\ref{our_recommendations}).}

\section{Background and Related Work}
%\subsection{Alexa}
\label{sec:background}

\subsection{{Alexa Ecosystem}} The Alexa ecosystem consists of a device placed in a user's home (e.g., smart speaker), a cloud platform implementing general voice assistant functionality, and an application endpoint implementing skill-specific functionality. The automatic speech recognizer (ASR) receives a command from the user and converts it to text. The natural language understanding (NLU) unit performs syntactic and semantic text analysis to determine the most appropriate \appv matching the user's utterance. The \appv developer hosts a cloud endpoint that implements the functionality API. For example, if the user utters \textit{``Lyft, Get me a car''}, the ASR and NLU will eventually determine that Lyft can handle the command. Thus, the \avS cloud platform sends a message to the HTTP(s) API endpoint for Lyft. 
Skill developers provide the directory entry information, and they are in complete control of all aspects of that data. This includes the skill name, invocation phrase, suggested commands and URLs for privacy policies and account linking.

%still needs a bit of work -- we need to discuss. The automating enabling is still not coming across. 
\subsection{{Skill Invocation}} \mjrev{
Amazon Alexa allows users to perform two skill operations - enabling and disabling. These operations can be executed via the skill's Amazon listing. Additionally, to promote usability Alexa automatically enables a skill when it is invoked by the user (e.g.,``Open Fitbit''). Once enabled, the skill can also be implicitly invoked (e.g., ``Ask Fitbit how I slept last night''). Lastly, a currently disabled skill cannot be invoked until it is explicitly re-enabled.}

% attacks on voice assistants
\subsection{Voice-based Confusion Attacks} 
\mjrev{
Prior work has demonstrated the existence of frequently occurring and predictable errors in Amazon Alexa's speech recognition engine. These errors can be leveraged to develop malicious skills (with identical or similar invocation phrases) that can hijack the voice command meant for a legitimate skill. For example, the voice command saying ``Alexa, open Fitbit,'' which is meant to invoke the \textit{Fitbit} skill, but can trigger the malicious skill \textit{Phitbit} after it has been published to the skill market. Upon activation, all voice interactions are handed over to the running skill which can perform a wide range of malicious activities like ask for private information or pretend to terminate and yet continue to operate by impersonating either other skills or the Alexa platform itself~\cite{zhang}.
}
Kumar et al. introduced skill squatting attacks, where a malicious developer registers a phonetically similar sounding skill as the target~\cite{kumar}. Zhang N. et al. perform a similar analysis for Google Home, and introduce an additional attack where a fake skill masquerades as a true one~\cite{zhang}. Finally, Zhang Y. et al. introduce \textit{lapsus} attacks that rely on common speech variation among humans~\cite{lipfuzzer}. For example, given a \appv with invocation \textit{``the true bank skill,''} a user might misspeak and instead ask for \textit{``the truth bank skill.''} An attacker can systematically discover common speech variations for a given phrase and then register skills. 
Broadly, all these attacks define a class of voice-based confusion attacks. The fundamental cause is the mismatch between a user's intention and the voice assistant's behavior.
%Our goal is to anticipate a user's intentions by analyzing contextual information from their counterpart apps and then enforce those intentions using a secure system design.

\subsection{Design Issues in Alexa Ecosystem} Lentzsch et al. recently performed a security analysis of the Alexa ecosystem that includes the vetting processes~\cite{alexa-skill-ecosystem-2021}. Like us, they find that attackers can circumvent the vetting and upload malicious skills to launch voice confusion attacks. They also observe that an attacker can mirror the metadata of legitimate skills, lending further confirmation that the ecosystem currently lacks any notion of secure identity. By contrast, our work contributes a method to provide a secure identity to skills through backlink search. 

% defenses against voice-based confusion attacks
\subsection{Existing Defenses} A direct way to combat voice-based confusion attacks is to prevent \appvs from being created with overlapping or identically-sounding invocation phrases. Kumar et al. suggest a phoneme-based analysis to detect similar-sounding skills ~\cite{kumar}. Amazon Alexa includes developer guidelines that explicitly prohibit duplicate names~\cite{alexa-guidelines}. However, these guidelines are not technically enforced. Recent changes to Alexa indicate that they now perform manual vetting, which is not scalable. Google Home's manual analysis detects when an invocation phrase is too close to an existing one. Although this seems like a reasonable approach, there are pitfalls. For instance, Google Home is prone to a phenomenon similar to domain squatting --- a third-party developer scoops up invocation phrases that heavily overlap existing and popular services (e.g., Lyft, Papa John's), preventing the first parties from creating \appvs with those invocation phrases~\cite{voicebot}. This also has the effect of third-parties benefiting from the copyright and reputation of the first-party invocation phrases on which they squat. %Furthermore, the lapsus attacks discussed above do not require the attacker to mirror the target invocation phrases completely. 
In summary, although these techniques add defense-in-depth, they do not address the root cause of the problem.

Zhang N. et al. suggest a skill response checker that keeps track of the current skill's responses and computes a similarity score to other skills~\cite{zhang}. If two skills are found to be similar enough, an alarm is generated. Our work is orthogonal and represents a systems-oriented defense that securely predicts and enables skills using counterpart app information. We observe that a secure voice assistant platform benefits from both approaches.

Guo et al. built SkillExplorer, an NLP-based testing system that uncovers potentially policy-violating behavior~\cite{skillex}. By contrast, our work ensures that only the user-intended skills run in response to ambiguous utterances and does not pass judgment on whether the authorized skill violates policies.

\subsection{Attacks on Speech Processing} A large body of recent work attacks voice assistants physically by injecting commands that exploit the ML models using adversarial examples~\cite{hvc,noodles,devil} or the non-linear components of microphones~\cite{dolphin,commander-song,inaudble} or the photoacoustic effect~\cite{sugawara2020light}. From \sys\xspace's perspective, these are simply voice commands that come from a non-human source. Our work is independent of the origin of a command, and it is not intended to address physical attacks. Instead, our work focuses on disambiguating confusing voice commands by analyzing counterpart activity and then ensuring that the disambiguated command executes.

%ensuring the voice assistant only executes legitimate skills that match the counterpart activity of a user. 

\section{Challenges in Preventing Voice Confusion Attacks}
\label{security_challenges}

\color{black}

% Although a key difference between voice assistants and other computing devices is the general lack of visual feedback, 
There are several fundamental design deficiencies that lead to voice-based confusion attacks. We identify them next and discuss how they guide the design of \sys.

%which we identify, in the ecosystem that lead to voice confusion attacks. We analyze these deficiencies and derive objectives that guide the design of \sys.
% also discuss why point-fixes do not sufficiently address the problem.

\subsection{Ambiguity in Speech Recognition}
Natural language ambiguity is an unavoidable source of confusion for voice assistants. Homophones, homonyms, and various pronunciation intricacies have the potential to produce an error in the ASR component of the voice assistant~\cite{kumar2018skill, lipfuzzer, zhang2019life, voicebot, zhang, flowfence16, skillex}. For example, ``Capital One'' and ``Capitol Won'' sound identical, and in the absence of other information, it is not possible to identify the user's intended skill. Improvements in ASR accuracy will not solve this problem. Alexa could present a choice to users when confused; however, as discussed next, Alexa lacks the notion of a secure identity that prevents users from identifying which skill is legitimate, given a choice. 

\noindent
\mjrev{\textit{\textbf{Objective 1:}} Seek a mechanism to resolve the ambiguity between phonetically similar skills that match a user's voice command.}

% To resolve natural language ambiguity, identify the skill a user wishes to interact with by

%Capture the intention of the user to identify the skill they wish to invoke as to help resolve the natural language ambiguity.
\begin{figure}[t]
    \centering
    \includegraphics[width=0.55\columnwidth]{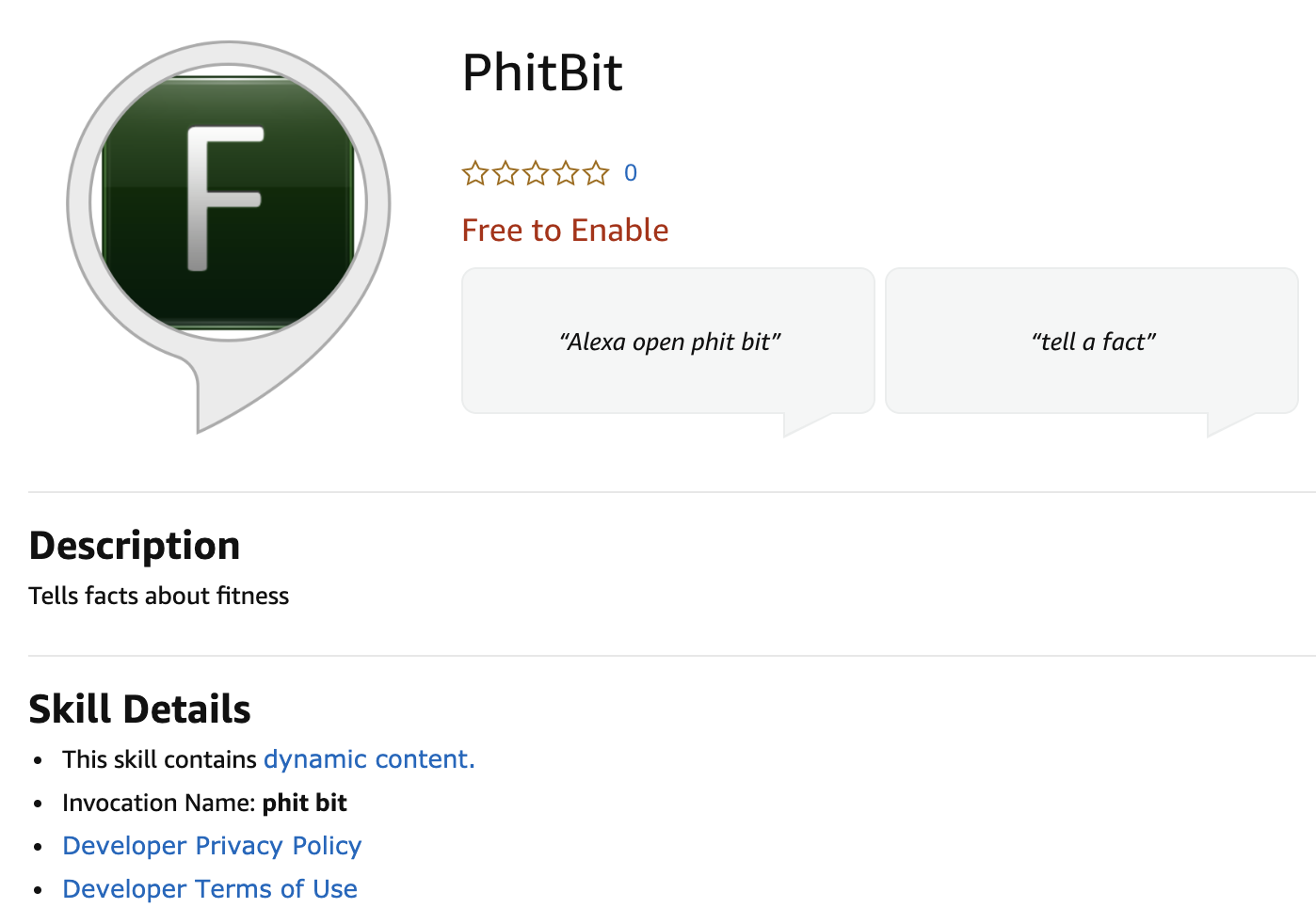}
    \caption{Metadata attack on FitBit skill. Our attack skill targets the true FitBit skill including its account linking URL.}
    \label{fig:phitbit}
\end{figure}

\subsection{Incompleteness of Metadata Vetting}\label{insufficient-vet}
Voice assistant ecosystems like Alexa vet the metadata of a skill before publication. This metadata includes the skill's name, account linking URL, privacy policy URL, developer URL, invocation phrase, and description. Recent work shows that an attacker can mirror the metadata of a legitimate skill and get it published to the marketplace~\cite{alexavetting, hu2020case}. To an end-user, two different skills might appear identical to each other. As Alexa skills do not have a notion of secure identity, attackers can impersonate legitimate skills.

Implementing stricter vetting policies does not address this problem completely. There are valid reasons for skills to share metadata, such as the invocation phrases. For example, our market-scale analysis of Alexa skills finds that \alexaSameName skills share their \emph{name} with at least one other skill and \alexaSameInvoke skills share their \emph{invocation phrase} with at least one other skill. If Amazon decides to prevent duplicate names, which of these skills to remove becomes unclear. Furthermore, requiring uniqueness in skill names can lead to a domain squatting effect --- a third-party developer scoops up invocation phrases that heavily overlap with existing and popular services (e.g., Lyft), preventing the first parties from creating skills with those invocation phrases~\cite{voicebot}. \mjrev{An alternative solution to this problem would involve prompting the users to decide which skills to invoke and then remembering their decisions. This approach requires user involvement (ideally through a screen), which goes against Amazon's rationale for automatically enabling skills.}

%  Alternatively, simply allow duplicate skill invocation phrases, but not skill names and icons -> seems like a middle ground when combined with prompt-then-remember
% Nov 5, 2021 3:13 PM

\subsubsection{Case Study}  We \mjrev{developed} the skill \emph{PhitBit} with the invocation phrase \texttt{`phit bit'} that sounds similar to Fitbit, a popular fitness skill handling sensitive information. Fitbit, being an \emph{account linking} skill, is required to provide an account linking URL in its metadata, which allows users to log-in and authorize it to access their Fitbit account. Figure~\ref{fig:phitbit} shows the Amazon Alexa skill page for PhitBit. \mjrev{When we submitted this skill for review,} Amazon's vetting process prevented us from copying Fitbit's graphics, but it did \emph{not} prevent us from using Fitbit's HTTPS API as the account linking URL nor Fitbit's privacy policy URL and developer URL. Our attack skill eventually passed the vetting process and was approved.\footnote{
%We ensured that our attack skill does not steal user data and we unpublished it after approval to ensure that real users do not invoke it.
\mjrev{Developers have two options while submitting a skill for review - (1) Certify and publish now, and (2) Certify now and publish later. We only submitted the custom skill for certification and never published the skill. Therefore, no users of Amazon Alexa interacted with the skill.}
}
Multiple submissions were made to get the attack skill, \emph{PhitBit}, approved. During one of the submissions, the vetting team informed us that the skill was not working because when they tried to invoke it, the command routed to the true FitBit skill instead of ours. On a subsequent submission, the voice command was routed to \emph{PhitBit}. This incident shows that the vetting process itself is vulnerable to voice confusion. \mjrev{The certification of this skill violated multiple security testing policies described by the Alexa skill review process. We disclosed this vulnerability to Amazon via the Amazon Vulnerability Research Program on HackerOne.}

% These flaws in the vetting policy allow any attacker to evade the vetting process, allowing them to publish skills that imitate existing ones. An attacker can publish multiple malicious skills with identical or similar invocation phrases to launch a voice confusion attack that is highly effective.

% \ah{The objective talks about TLS, but this hasn't been discussed in the text above}

\noindent
\textit{\textbf{Objective 2:}} Establish the secure identity of a skill to prevent confusion with other skills.

\subsection{Lack of Runtime Security Indicators}
Once a malicious skill is invoked, it is easy for the attacker to deceive end-users --- fake skill interactions mimic either the Alexa voice system or that of any other skill~\cite{major2019alexa, alexa-no-perm-prompt, voicebot}. Most Alexa users are unable to differentiate between genuine and malicious skill interactions~\cite{major2019alexa}. Unlike visual interactions, communicating security information over voice is not straightforward; it is hard to incorporate an HTTPS-like icon in a voice interaction. Once executed, a fake skill can ask the user for information that the user only shares with the target skill.  For example, Find My Phone asks for a user's phone number, Transit Helper requests a home address, and Daily Cutiemails solicits a user's email address. This kind of attack can also damage the reputation of a legitimate skill, given that any poor service of the malicious skill will be blamed on the legitimate skill. Finally, the fake skill can lead to a phishing attack by delivering misleading information, such as a fake customer contact number, through the voice channel to the user. Thus, any defense strategy that involves the user interacting with a skill (after its execution) is insufficient. 

\noindent
\textit{\textbf{Objective 3:}} Make the security decision (about the legitimacy of the skill) \emph{before} a skill runs; only user-intended skills should execute.

\subsection{Closed Skill Ecosystem}
The code for an Alexa skill executes on the third-party developer's backend service that they fully control. Different from other ecosystems (e.g., Android), the skill code is not available to end-users or even Amazon. Thus, solutions based on OS changes to the Alexa device or code analysis techniques on the skill are not possible. Furthermore, the Alexa skill ecosystem is closed to incorporating built-in defense strategies from third parties, unlike the Android or web ecosystems.

\noindent
\textit{\textbf{Objective 4:}} Develop the defense strategy to be practical for deployment without modifications to Alexa.

%%%\sys is the system name. 

\section{\sys\xspace Design}
\label{sec:design}

In the following, we describe the design of \sys, starting with the system and threat model, followed by an operational overview. We then elaborate on the individual components of \sys.

\subsection{System and Threat Model} We assume a malicious skill developer that launches voice-based confusion attacks~\cite{kumar,zhang,lipfuzzer}. These attacks depend on the user's inability to identify the precise skill they are interacting with and on errors in the speech processing components.  The attacker's goal is to get their malicious skill executed so that they can use it to conduct phishing or execute other restricted actions~\cite{kumar,zhang,lipfuzzer}. As the current \avS model does not properly vet skill metadata (Section~\ref{insufficient-vet}), an adversary can upload malicious skills with arbitrary metadata, including metadata that is directly copied from legitimate skills. We assume that the end user's devices are trustworthy: the voice assistant itself, smartphones, and web browsers. We also assume that the websites of the legitimate skill developers are not compromised. For example, the Fitbit or the NBC website is not compromised and is trustworthy.

\subsection{Operational Overview}

\mjrev{
Skill invocations are not equally ambiguous.  For example - the ``Fitbit'' skill is much more likely to be \emph{confused} with the ``Phitbit'' skill rather than the ``Lyft'' skill. For skills that are phonetically ambiguous, \sys uses information from the user's history to nudge the \avS towards executing the skill that matches the user's preferences --- thereby resolving the confusion. There are four questions associated with \sys's operation: (1) \textit{how to identify skills that are phonetically similar?} (2) \textit{how to identify the user's preferences?} (3) \textit{how to match the user's preferences to a skill?} and (4) \textit{how to nudge the \avS to execute the matched skill?} In the following, we discuss -- at a high level -- how \sys's design answers these questions while achieving the four objectives from the prior section.}

\subsubsection{Phonetic Similarity} \mjrev{We model the phonetic similarity relationship of skills as a phonetic graph. In this graph, skills are the vertices, and the edges represent skills that are phonetically similar, i.e., potential targets for voice confusion. \sys uses this graph to reduce the potential of voice confusion attacks. }

\subsubsection{Collecting the User's Preferences} \mjrev{An end-user installs the browser extension that we have built, and it periodically processes counterpart information --- browser history and Android apps (that the extension collects by navigating to the \url{play.google.com} website). This counterpart information represents the user's preferences (\textbf{objective~1}). For example, assume the user has purchased a Fitbit device. It is likely that they will have accessed Fitbit websites or installed the corresponding Android app. Later on, they might start using the Fitbit Alexa skill. The \sys\xspace extension processes the prior URL visits or Android app installation information and encodes this information in the user's phonetic graph to resolve the confusion at run-time. Thus, when the user speaks the command ``talk to Fitbit,'' \sys\xspace will know they mean the fitness tracking skill and not some other neighboring skill in the phonetic graph.  We note that browser history and android app collection occur in a privacy-respecting way. This data is never transmitted outside the extension; rather, matching browser history to skills occurs locally. Furthermore, we implement existing techniques to ensure that our system is not tricked by web history poisoning (see Section~\ref{subsubsec:adaptive-attacker} for more details).}

\subsubsection{Securely Mapping Preferences to Skills} Matching the counterpart activity to the Alexa skill requires a secure skill identity (\textbf{objective~2}). However, as we have discussed, the Alexa ecosystem does not have such an identifier. A contribution of our work is a practical way to achieve secure identity by only requiring small changes on the skill developer's websites. For a given skill, the \sys\xspace backend extracts all URLs from metadata (e.g., account linking URL, privacy policy URL, developer URL). It then searches through these websites for a link \emph{back} to the Amazon Alexa marketplace listing for the skill. If such a backlink exists, then we associate the identity of that website (i.e., HTTPS certificate) with the skill. In our running example, \sys maps the user's visits to \url{fitbit.com} to the correct Fitbit skill.

\subsubsection{Practically Ensuring that Only Matched Skills are Executed} \mjrev{ 
After \sys\xspace has identified and securely matched the user's preferences to a skill, it must ensure that \avS executes the matched skills, despite confusing voice commands and the presence of possible attacker skills (\textbf{objective~3}). In our example, \sys must ensure that the Fitbit skill executes instead of the Phitbit skill. Furthermore, \sys\xspace must achieve this task on the existing Alexa system without modifications (\textbf{objective~4}). Our work overcomes this challenge by discovering and characterizing a feature in Alexa skill execution: given two identical skills, if one of them is in the \emph{enabled} state while the other is in the \emph{disabled} state, Alexa always executes the enabled one. \sys leverages this behavior to explicitly enable the matched skills and disable their neighboring skills -- nudging Alexa towards executing the matched skill. Back to our example, \sys enables the correct Fitbit skill and disables the Phitbit skill (as well as the other neighboring skills). When the user speaks the command ``open Fitbit,'' Alexa will not be confused with the Phitbit skill and will execute the correct Fitbit skill. Section~\ref{backend} elaborates more on this aspect, which we empirically characterize in Section~\ref{sec:evaluation} using large-scale voice experiments.}

\begin{figure}[t]
  \centering
  \includegraphics[width=0.7\columnwidth]{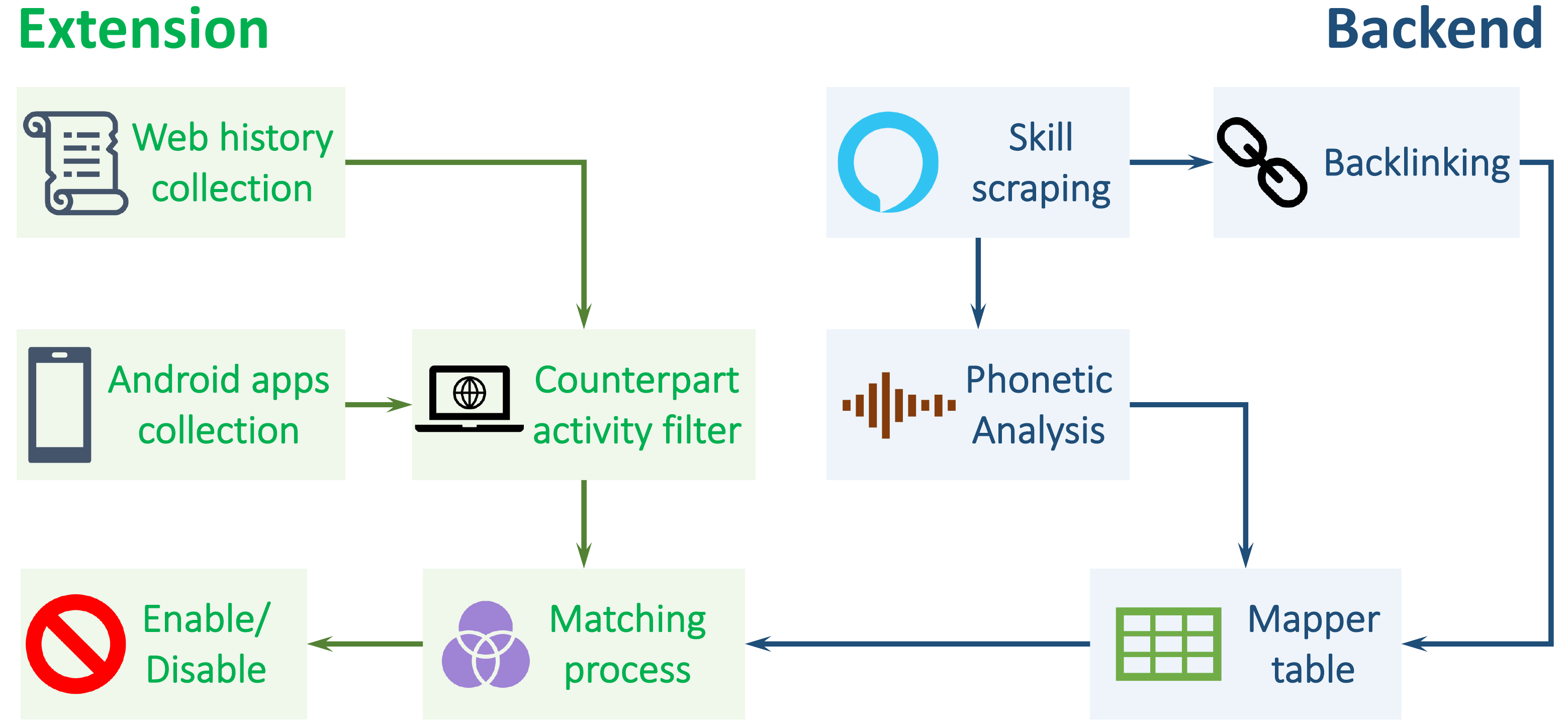}
  \caption{End-to-end system overview diagram.}
  \label{fig:high_level_diagram}

\end{figure}

\subsubsection{Putting Everything Together} \mjrev{
\sys\xspace consists of a user-facing extension and a backend component as shown in Fig.~\ref{fig:high_level_diagram}
The backend component of \sys periodically scrapes the Alexa skill metadata. It performs backlink analysis to establish a secure skill identity. It also models the phonetic similarity among skills as a phonetic graph. \sys's extension periodically collects the user's preferences from counterpart activity and securely maps them to a set of skills. It then uses the phonetic similarity graph to enable/disable the appropriate skills to reduce voice confusion attacks. In particular, \sys's extension enables skills that match the user's preferences from counterpart activity and disables their neighbors. The design of \sys\xspace achieves \textbf{objective~4} as it requires no change to either the skills or the Alexa ecosystem; it leverages existing tools to meet the first three objectives. In the next two sections, we discuss each of the above steps in detail, outlining the underlying algorithms and data structures.
}

\subsection{Backend Component}
\label{backend}

The backend component of \sys periodically analyzes the skills in the Alexa marketplace. It performs two operations: (1) secure identity generation using backlink analysis and (2) phonetic graph construction for modelling skill phonetic similarity (which the extension eventually uses to enable/disable skills). The backend distills the results of these two operations into a \emph{mapper table}.

\subsubsection{Secure Identity Generation}
\label{secure_backlinking}

\begin{figure}[t]
    %\centering
    \includegraphics[width=0.7\columnwidth]{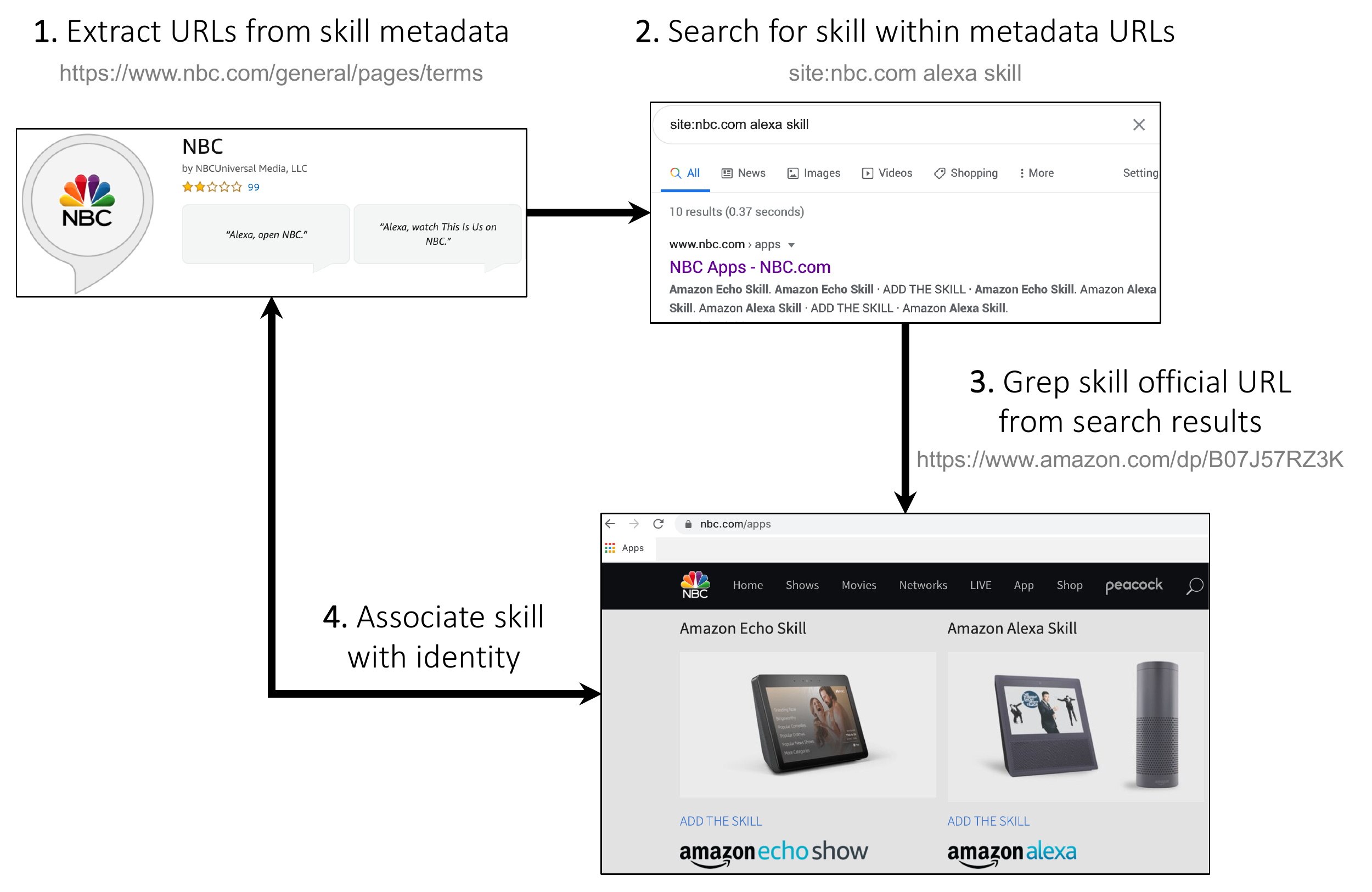}
    \caption{Generating secure skill identity: \sys  searches for backlinks to the same skill's Amazon listing within domains that are extracted from the skill's metadata. If there is a link from a domain back to the skill, then the skill-domain pair is added to the mapper table.}
    \label{fig:secure_identity}
\end{figure}

The key insight for generating secure identity is to leverage trust in the existing PKI and web ecosystem. \sys searches through the domains in a skill's metadata for a URL linking back into the Amazon marketplace listing for that skill. Assuming that the malicious-skill attacker in our threat model has not also compromised the legitimate skill's website, the presence of the \emph{backlink} indicates that indeed, the developer of the website and of the skill are the same entity, and thus, the identity of the skill is the identity of the website (i.e., the TLS certificate). Figure~\ref{fig:secure_identity} illustrates the high-level idea, for the NBC skill as an example. \sys backend will extract all metadata URLs and search for a URL to the NBC Alexa skill page. Once found, it will associate NBC's website identity with the skill in the mapper table data structure.

Concretely, the backend extracts URLs from each skill's metadata, including the privacy policy, the developer website, and the account linking URLs. Then, it extracts the root domain of each URL to search for webpages that link to the skill. In particular, \sys issues Google search queries of the type - site: $<$domain$>$ ``alexa skill." Then, it uses Selenium to crawl the webpages for the top 100 search results, while searching for the skill's Amazon URL. When it finds such a link, \sys establishes a mapping between the public key certificate (if available) of the domain and the skill. \sys also uses link backtracking services~\cite{Gerasimenko:2010} to find whether any of the websites from the skill's metadata link to the skill's Amazon page. 

\color{black}

\mjrev{This method presents exhibits two advantages. First, because it is automated, it scales over a large number of skills in marketplace. Second, it allows skill developers control over improving security for their skills rather than relying on Amazon’s vetting process. Compared to alternatives, such as updating the voice assistant platform to support cryptographic signatures,} this design requires a simple change on skill developer websites; they need to only include a backlink to the Amazon marketplace listing of the skill. \mjrev{This also serves to improve the credibility of the skill as Android users too are recommended to rely on official websites of popular services like Facebook, WhatsApp, Truecaller to find trusted sources for links to the apps ~\cite{android_backlink}}. We have begun explaining the benefits of this approach to skill developers and convinced one developer to update their website with a link to their skill (\emph{OurGroceries}\footnote{Skill - https://www.amazon.com/HeadCode-OurGroceries/dp/B01D4F1J0M \\ Website - https://www.ourgroceries.com/user-guide\#installing}), demonstrating the value of our approach. An entity like Amazon can easily require skill developers to implement this change.

%We note that ideally, Amazon could also require skill developers to directly include the backlink URL directly in skill metadata. This would remove the need for a backlink search process. However, until voice assistant vendors decide to adopt our changes, users still require protection. \sys protects users today by using the backlink search to automatically generate secure skill identities.

%\color{blue}
%\noindent{\textbf{Trade-off:}} While the backlinking process provides secure mapping, it would not be able to consider websites that currently do not have backlinks in place. An alternate mapping method could be to check for consistency in skill metadata between the Amazon homepage and the developer website. While this method may have a wider coverage, it is susceptible to attacks. For example, a malicious skill may mimic the metadata of a popular skill, including the linked URLs.
%\color{black}

\subsubsection{Phonetic Graph}
\label{sec:phonetic-graph}
%% why do we need the phonetic graph?
\mjrev{
\sys uses the phonetic graph to identify potential targets of voice confusion. To construct the graph, we first model the phonetic similarity between skills using a distance metric. This metric is used to derive a weighted graph where the vertices are skills and edges are weighted by the phonetic distance. \sys uses this graph to generate a list of phonetic neighbors for each skill, which are added to the mapper table.}
%The \sys extension must efficiently disable skills that do not get mapped to any counterpart activity. As the set of unmapped skills is always much larger than the set of mapped skills, disabling them becomes very inefficient because of Alexa's restrictive API that only permits disabling or enabling single skills at a time. Our insight is to analyze the set of skills in a phonetic radius around a mapped skill and disable ones within the radius because those skills are most likely to cause a voice confusion attack.  We represent all the skills in the Alexa market as a fully connected and undirected graph. The vertices are the individual skills, and the edge between a pair of skills is weighted by the phonetic distance between their invocations.  

%As discussed earlier, this graph helps the enable/disable module of extension to disable the phonetic neighbors of the user's enabled skills.

\paragraph{Phonetic Distance.}
\label{ph_dist}
We define a distance metric that quantifies the confusion between two skills based on the phonetic representations of their invocation phrases.  This metric ensures that different phoneme transformations have different costs. To achieve this property, we compute a weighted Levenshtein distance between the phonetic representations where different phoneme edit operations have different costs. We compute the weighted cost matrix by applying the Needleman-Wunsch Algorithm on the CMU dictionary, which has around  $9181$ pairs of alternate pronunciations (also used in prior work~\cite{zhang}).
For example, the substitution cost for replacing phoneme ${\alpha}$ with phoneme ${\beta}$:
\begin{equation*}
    WC(\alpha, \beta) = 1 - \frac{SF(\alpha, \beta) + SF(\alpha, \beta)}{F(\alpha) + F(\beta)},
\end{equation*}
where $F(\alpha)$ is the frequency of occurrence of phoneme $\alpha$ and $SF(\alpha, \beta)$ is frequency of occurrence where phoneme $\alpha$ has been substituted by phoneme $\beta$ in all alternate pronunciation pairs in the corpus. This cost function ensures that phoneme substitutions that provide valid alternate pronunciations of English words are given a smaller weight. We also normalize this distance by the length of the invocation phrases.

\begin{figure}[t]
    \centering
    \includegraphics[width=0.6\columnwidth]{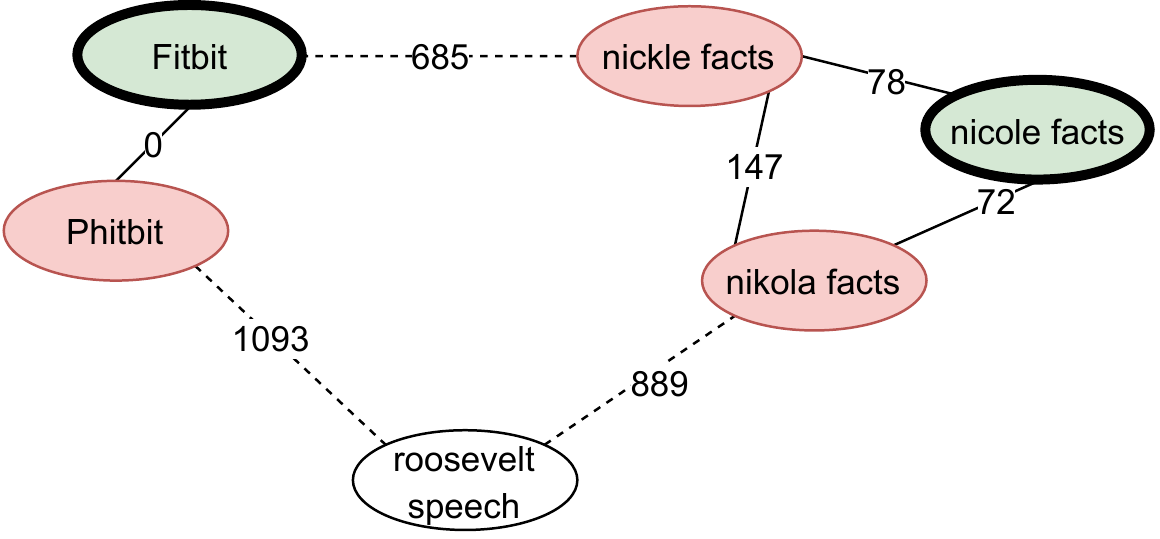}
    \caption{Phonetic graph of a set of skills with phonetic distance between their invocation phrases as weights. The skills ("nicole facts", "Fitbit") are mapped using counterpart activity. Note: the dashed lines are dropped edges. Green/Red nodes represent skills that are enabled/disabled by \sys respectively. }
    \label{fig:skillgraph}
\end{figure}

\paragraph{Constructing the Phonetic Graph.}
\label{phonetic_similarity}
\mjrev{
We utilize the phonetic distance to compute the weights of the edges in the phonetic graph. We then prune the graph by dropping all the edges with distance greater than a threshold; the tuning procedure  of the threshold is discussed in Section~\ref{sec:evaluation}. The edges in the resulting phonetic graph represent instances that are prone to confusion attacks. The skills with no neighbors are less prone to speech interpretation errors. As an illustration, Figure~\ref{fig:skillgraph} shows the graph for six skills, represented by their invocation phrases. For a phonetic distance threshold of $200$, the graph has only four edges (represented as solid lines). In this example, the nodes with thick borders represent the skills ("nicole facts", "Fitbit") matched using counterpart activity collection. The extension component enables the matched skills (colored in green) and disables their neighbors (colored in red). As we explain later, this process reduces the number of incorrect invocations  due to voice confusion (e.g., between ``Fitbit'' and ``Phitbit''). Skills not connected to an enabled skill need not be disabled. We evaluate the effectiveness of this approach in Section~\ref{eval_enable_disable}.}

\paragraph{Trade-off.}
\mjrev{
\sys aims to disable all skills connected to an enabled skill in the pruned phonetic graph. A smaller distance threshold would lead to fewer edges, resulting in an efficient disabling process. However, it increases the chances of incorrect invocation between unconnected skills. A sufficiently large threshold can reduce confusion, at the cost of having to disable a larger number of skills and thus, a less efficient disabling procedure. It also can increase the potential of disabling skills that the user might later need. We empirically tune the threshold to obtain a reasonable trade-off between potential for invocation confusion and efficiency of disabling at runtime (Section~\ref{sec:tuning}).
}
%The latter case, however, implies that \sys can enable only a smaller set of skills, which comes at a usability cost to the users. Hence, we tune the threshold such that we prevent cross-component confusion.

%ensuring a high number of components. %Can we clarify this?

\subsubsection{Mapper Table}
\label{mapper_table}
\mjrev{
Each row of the mapper table contains an Amazon marketplace skill URL, the domain and its certificate that forms the secure identity, and a list of neighbor skills (derived from the phonetic graph). The extension uses this information at runtime to index and enable/disable skills. The backend process continually updates the mapper table and the extension periodically downloads it (approx. 5 MB for the initial download).}

%The mapper table stores the counterpart app identifiers for every skill in the market, if applicable. It also contains phonetic distance information for each skill in relation to each other skill in the Alexa Marketplace; a data field for a skill being a singleton is also present so that such skills can be quickly identified for disabling.
% The mapper table also contains other metadata information that is referenced by the extension to enable the matching \appvs automatically during online execution. This table is continuously updated at the \sys  extension; it is roughly $5$MB in size.
%\mnotes{fill in the info about the mapper table...}

\subsection{Extension Component}
The browser extension of \sys is its user-facing component. It interacts periodically with the backend component to fetch the up-to-date mapper table. Also, it \textit{locally} extracts the user's activity with counterpart apps to enable the skills that match the user's activity. Finally, it uses the mapper table to identify and disable the skills within a phonetic distance from the enabled skills.

\subsubsection{Counterpart Activity Collection}
\label{site_history_filter}

\mjrev{
\sys runs as a Chrome-based browser extension. The extension periodically (every $5$ minutes) retrieves user's installed app data and browser history from their Google account. It scrapes the user’s “My Android apps” page on play.google.com/apps to obtain the list of installed Android apps. For browsing history, it uses the \textit{chrome.history} API to query the browser records. The extension only searches for URLs whose root domains are part of the Mapper Table. There are three considerations associated with this process: execution overhead, user privacy, and the potential for adaptive attacks.}

First, the extension periodically scrapes the installed apps to identify the newly installed apps. Further, it collects the history using browser APIs, which do not incur any noticeable overhead (<$5$ sec). Second, this process takes place \textit{locally}, and the collected information never leaves the user's browser. Third, an attacker might exploit this process to poison the user's browsing history. Without special attention to this case, \sys  could potentially use visits to attacker-controlled websites to enable attacker-controlled skills.   

\sys  counters the web poisoning attacks through a web filtering heuristic.
It only considers websites with the following properties: (1) the website is visited across multiple valid sessions, (2) each valid session consists of visits to at least three different pages within the website's domain, (3) the considered visits are not directly preceded by visits to skill pages (these criterion are based on past work)\cite{4221889, 4061378, history-filter}. The first two criteria focus on websites that the users are likely to visit without attacker influence through phishing or spamming. The third criterion prevents \sys  from recording visits to a website that are followed by visits to the skill pages.

\paragraph{Trade-off.} The selectivity of the web history filter can be adjusted to address the trade-off between usability and security. A more selective filter makes it harder for the attacker to get a malicious skill enabled, but it would also lead to the collection process missing out on genuine websites that the user actually visited. Whereas, a filter that considers most of the visits is likely to have a large coverage but would be easier to evade.

\subsubsection{Matching Process}
\label{site_to_skill_mapping}
\mjrev{
After URL filtering,  \sys matches the user's visited websites or installed apps to Alexa skills. \sys obtains the list of installed apps and the user's browser history. It then extracts the certificate from each visited website and installed app. Using the pre-fetched mapper table and these certificates, it matches the visited websites and installed apps to their corresponding skills. Recall that the mapper table stores associated certificates for each skill. This process leaks no information about the user's history of activity with counterpart apps. The matching is performed locally with the entries of the mapper table.}

\subsubsection{Enable/Disable Module}
\mjrev{
The final step is to enable the matched skills and disable their phonetic neighbors. Amazon's skill listing interface allows a user to perform two basic operations:  \textit{enable} and \textit{disable} skills. We empirically observe (Section~\ref{sec:evaluation}) that enabling a skill and disabling others with similar invocation phrases ensures that Alexa invokes the enabled skill and thus, prevents confusion attacks.}

\mjrev{\sys leverages this observation to enable all skills that were matched using the mapper table and user's history. Next, it disables the phonetic neighbors of the skills enabled in the previous step. The extension performs these operations on behalf of the user. Since the disabled skills cannot be invoked without explicitly enabling them first, this also acts a warning mechanism for incorrect invocation. }

% \mjrev{
% \sys performs the following operations for each mapped skill. First, it enables the skill if it is currently in the Disabled or Untouched state. Second, it disables all phonetic neighbors that are in the Untouched state. }

\subsubsection{Online Operation}
\label{online_op}
As discussed earlier, \sys is a Chrome browser extension. To run it, the user needs to: (1) install the extension on their Chrome browser; (2) allow the extension to access their web history and/or installed apps; and (3) allow the extension to enable/disable skills. If the user does not provide the required permissions, then \sys will not be operational. 

At installation time and each hour thereafter, \sys\space communicates with its backend server to fetch the latest mapper table. During its initialization on a new device, \sys fetches the the complete mapper-table of size $5$MB. Subsequent transmissions include delta updates (~$2$KB per skill). Assuming updates to around $26$ skills everyday~\cite{skill_stats}, this results in a total network overhead of 50KB/day. \sys retrieves the counterpart data and uses the latest mapper table to identify the matched skills.  
Then, \sys applies the Enable/Disable module for the matched skills. 

%  When the user utters an invocation, Alexa will always execute an enabled skill despite the presence of phonetically-close skills that \sys has disabled (Sec.~\ref{eval_enable_disable}). If the user tries to invoke a disabled skill, Alexa will show a permission prompt asking them to confirm that action.
 
\sys does not maintain per-device or per-account history. Given counterpart data, it enables/disables skills based on their current state in the user's Alexa account and does not interfere with previously-enabled skills. The counterpart data can come from different sources, such as web history, app installations on the same or different devices. Thus, the user can install \sys on multiple devices, and even configure different google accounts (for counterpart activity) against a single Amazon Alexa account. Despite this diversity of counterpart data sources, \sys operates with a single Alexa account at any given moment, and thus, it has an accumulative effect on that Alexa account even if the user has multiple computers (e.g., desktops, laptops, smartphones). %Therefore, \sys does not need to store per-device or per-account history 

%\sys independently retrieves the user's counterpart data from the devices and enables/disables the skills on the same Alexa account. Thus, it does not need to synchronize data across multiple devices.  

\mjrev{
Finally, \sys might not pre-enable some legitimate skills that the user wants to be executed. This could happen because there is not enough counterpart activity or the skill does not have a secure backlink. In such a case, \sys does not affect the user experience, it defaults to the state of the world today: Alexa will pick a skill according to its matching algorithm. As such, \sys does not deteriorate the security of Alexa users, but might harm their user experience when it disables skills they might need in the future. We discuss these trade-offs in the evaluation.}
%In such a case, there are three possible outcomes to the user experience:
%\begin{itemize}
%    \item The legitimate skill is naturally isolated in a phonetic island (no potential for squatting based on the current state of the skill marketplace).
    %\sys does not disable skills unless they lie in a phonetic neighborhood around a securely-predicted one.
%    In this case, there is insufficient data for \sys to predict the skill and thus, it takes no action. Alexa will just execute the skill which has a minimal squatting potential.
    %Alexa will just execute the skill, which  implies no usability burden and minimized squatting potential. 
    
%    \item The skill is in the phonetic neighborhood of another skill that is pre-enabled (because \sys had associated counterpart activity). In this case, Alexa will default to the pre-enabled skill (which is phonetically similar). This case (which is rare according to our evaluation in Sec.~\ref{end_to_end_effectiveness}) entails a usability limitation. 

%    \item The skill is phonetically similar to a set of skills, and none of them are securely predicted by \sys (because there is insufficient counterpart activity). In this case, \sys defaults to the state of the world today: Alexa will pick a skill according to its matching algorithm and squatting can occur. \sys does not affect the user experience. However, this case highlights the limitations of of Alexa's natural language understanding, which we evaluate in Sec.~\ref{eval_enable_disable}.
%\end{itemize}
% 
% 
% 
% 

\color{black}

\section{Evaluation}
\label{sec:evaluation}

We perform an end-to-end evaluation of \sys and it's core components. We summarize our evaluation findings, then discuss the experimental setup and finally provide experimental details. We also discuss how \sys's design protects against an adaptive attacker with knowledge of its inner workings.

\subsection{Evaluation Overview}
 
 Our evaluation answers the following questions:
\begin{itemize}

    \item[\textbf{Q1.}] \textbf{\textit{What are the characteristics of Alexa Enable/Disable API in controlling which skills get executed?}}
         We perform a large scale measurement to test \sys's Enable/Disable module against phonetically similar skills available in the Amazon Marketplace. These skills include ones that share invocation phrases or have similar sounding invocations. We establish that when a skill is enabled and its phonetic-neighbors are disabled, Alexa always executes the enabled skill.

    % \sys is able to bring down the rate of incorrect invocations from 27\% to 0\% for skills with similar sounding invocations.

     %--end to end attack reduction
     %--usability-security trade-off - FAR, FRR
     %--setup time
    \item[\textbf{Q2.}] \textbf{\textit{How effective is \sys in preventing voice confusion attacks? How does \sys balance the trade-off between security, usability, and performance?}}\\
         \mjrev{We perform a trace-based evaluation with real user data from our data collection effort. For a distance threshold of $400$, \sys is able to perform at a False Rejection Rate of $9.17$\%, a False Acceptance Rate of $19.83$\% and an Equal Error Rate of $16.5$\%.
         Our evaluation also shows that the initial setup time depends on the phonetic distance threshold. The optimal threshold between usability and security corresponds to one hour of initial setup time. This initial time is a side-effect of implementing \sys with a closed system; a better API design from the manufacturer can significantly cut this time down. Subsequent \sys operations encounter fewer enabled skills at a time, leading to a much quicker response time of less than a minute.}

    \item[\textbf{Q3.}] \textbf{\textit{How does \sys protect against an adaptive adversary? What are the privacy implications of using \sys?}}\\
        The different components of \sys are designed to address an adaptive attacker and make it more challenging to launch a successful skill-squatting attack. \sys does not maintain any state or store any user information, even locally, thus minimizing any privacy concerns.
    
    \color{black}
    
\end{itemize}

\subsection{Experimental Setup}
\mjrev{First, we perform a large-scale measurement to characterize the phonetic similarity thresholds of Alexa and confirm its enable/disable behavior. Second, we perform trace-based evaluation to study the user experience of a \sys user. We collect the counterpart history and the enabled skills from a set of Alexa users. We use the collected data to mimic the user experience with \sys. First, we initialize the backend of \sys by populating the Mapper Table. Then, we use the counterpart history data of each user to enable/disable skills on their phonetic graph. We create a test Amazon account for each user and populate it with their phonetic graph; we then invoke skills from the user's reported history as well as their phonetic neighbors. Effectively, we characterize the security gains as well as the usability and performance overhead for each user.}

We focus our analysis on a subset of the Amazon Alexa skill Marketplace: the \acLinkingSkills skills with account linking, which represent sensitive skills on the Alexa market according to a recent analysis by Shezan et al.~\cite{shezan-www2020}. Account linking skills are likely to be the targets of attack because of their access to the user's account data. We designate skills that have account linking as critical. Note that \sys's design, implementation, and evaluation can be readily applied for the full set of skills without changes to the Alexa Marketplace or how skills operate on the Alexa platform. 

\subsubsection{SkillFence Backend}

 \begin{figure}[t]
    \centering
    \includegraphics[width=0.5\columnwidth]{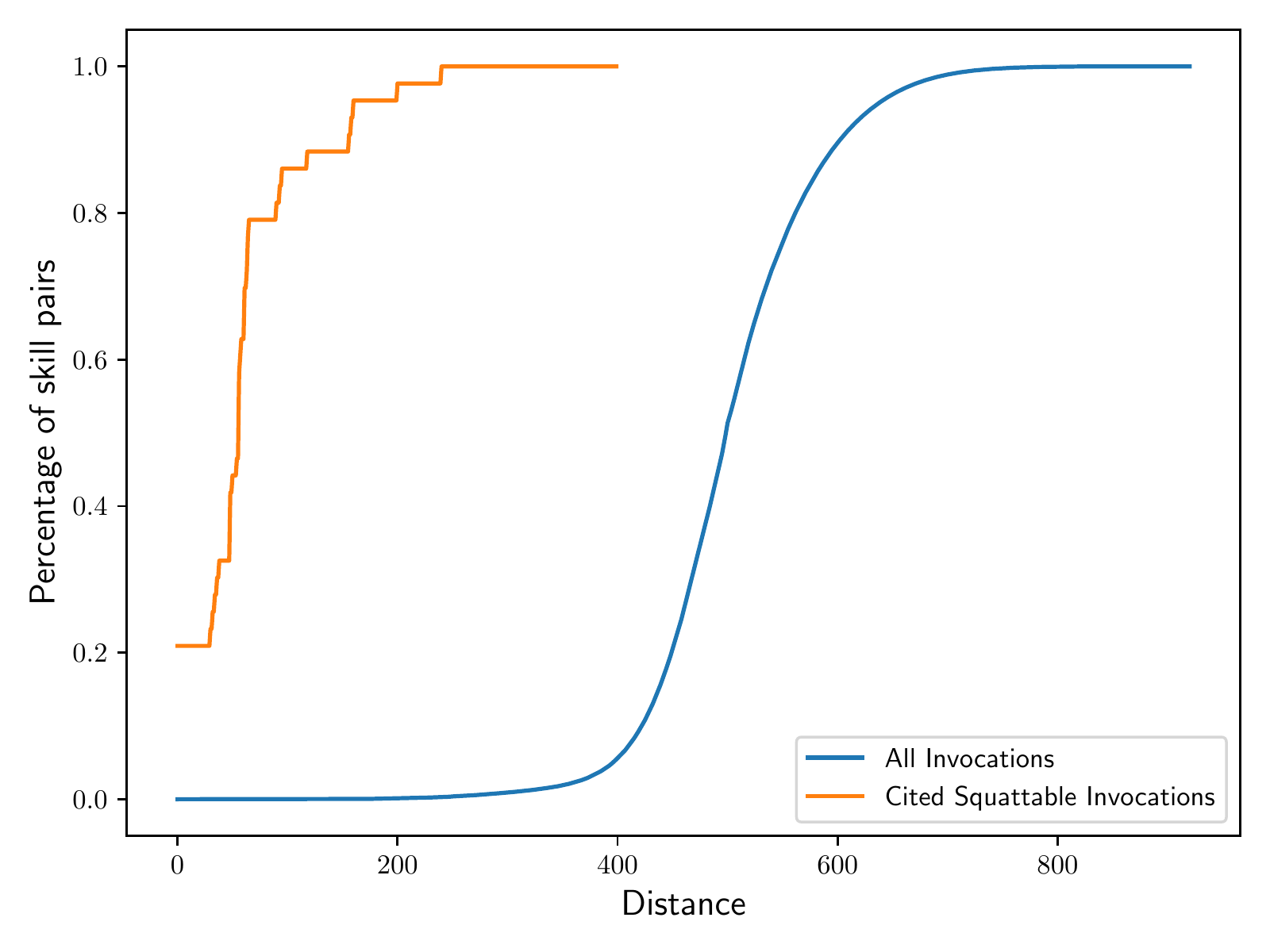}
    %\label{fig:dislit}
    \caption{CDF of phonetic distance between invocations of all possible Skill pairs. More than $95$\% skill pairs have phonetic distance larger than $400$. All skill squatting examples cited in previous work lie within a phonetic distance of $250$.}
    \label{fig:dislit}
\end{figure}

% \noindent\textbf{Mapper Table}\label{sec:mapper-table}
We generate the Mapper Table for the \acLinkingSkills Amazon Alexa skills with account linking. This process includes running the backlinking analysis to extract the secure links of the skills and constructing the phonetic graph. 

\paragraph{Backlinking}
The majority ($2496$ out of \acLinkingSkills) of skills contain at least one valid unique domain in their provided metadata. Skills that share domains generally belong to the same developer \mjrev{(we do not consider domains of popular cloud hosting services such as Amazon and Google, for backlinking)}. A skill-domain pair is part of the Mapper Table only if a website associated with the domain lists a link to the skill's Amazon homepage. To search for backlinks, we follow two approaches -- Google Crawling and Link Backtracking Services. For Google Crawling, we constructed queries like "site:$<$domain name$>$ Amazon Alexa Skill" and searched through the top $100$ results. We also used additional text phrases such as `Alexa Skill,' and `Amazon Skill,' to exhaustively search for all Alexa skill references. These queries provided at least one search result for $2023$ skills. Among these, we found backlinks for $404$ skills. For building a more exhaustive backlink profile, we utilize a Search Engine Optimization (SEO) service, \textit{Ahref}\footnote{\url{https://ahrefs.com/backlink-checker}}, which provides a tool for retrieving all backlink references for a given URL. Overall, employing both approaches, we were able to find the backlinks for $474$ skills. This number is likely to increase as websites start adding references to their skill's homepage like they currently do for their Android and iOS apps. \mjrev{Android users are recommended to rely on official websites of popular services like Facebook, WhatsApp, Truecaller to find trusted sources for links to the apps ~\cite{android_backlink}. The secure identity generation process automates this process on the user's behalf. Furthermore, as we discuss later, we have begun outreach efforts to encourage skill developers to include a link to their Alexa skills on their webpage, and we are beginning to see adoption.}

% We found that 1452 out of \acLinkingSkills skills have at least one backlink. 
\paragraph{Phonetic Graph}
We constructed the skill phonetic graph consisting of \totalSkills skills. Figure~\ref{fig:dislit} shows a CDF of weight values of edges in the Phonetic Graph, i.e., the phonetic distance between all pairs among \totalSkills skills. The mean distance is around $500$ and most of the distance values are less than $800$. We also observe that only $5$\% of the weights lie below the value of $400$. This suggests that pruning all edges with weights more than $400$ would lead to a sparse phonetic graph and considerably reduces the number of skills to be disabled. This is important because it improves \sys's efficiency. Figure~\ref{fig:dislit} also shows that all skill squatting invocation pairs introduced in prior work have phonetic distances less than $400$~\cite{zhang,kumar}. 
%On the other hand, we observe no incorrect invocations for skills pairs with distance greater than or equal to 500.  Hence, there exists a trade-off between restricting the number of skills to be disabled and identifying potential skill squatting pairs.

\subsubsection{Data Collection}
\label{data_collection}
We performed a data collection campaign to collect installed mobile apps, skill-relevant browsing history, and preferred Alexa skills from a set of individuals through Amazon Mechanical Turk (MTurk). This data allows us to perform trace-based evaluation of \sys's extension component. 
We focused the data collection on individuals who own Alexa devices and Android smartphones. We obtained IRB approval for the data collection procedure from our institution. 

\paragraph{Data Collection Design}
We designed a Qualtrics survey hosted on MTurk to guide individuals through the data collection procedure. The survey consists of five sections: eligibility, Chrome extension installation, data collection, data validation, and clean-up. After completing a consent form that verifies age, Alexa ownership, and records the MTurk ID, the participants install a Chrome extension and log into their Amazon and Google accounts. The users enter their MTurk IDs again in the extension as a verification check. 

The data collection procedure involved two components: (1) extracting user data consisting of enabled skills, browsing history, and installed Android applications; and (2) validating this information through a survey form. We created a Chrome extension to automate data extraction. This extension scrapes the enabled Alexa Skills from a user's  ``Your Skills" page on \textit{alexa.amazon.com}. It also scrapes the user's ``My Android apps'' page on \emph{play.google.com/apps} to obtain the list of installed Android apps. In the browsing history, the extension only searches for URLs whose root domains are part of the Mapper Table --- other URL data is not transmitted to our servers.

There are two potential problems with the extracted data. First, the user could have cleared their browsing history. Second, a user might have enabled a skill with which they do not interact. Note that the first issue applies only for the data collection and does not affect \sys  as it runs continuously at the user's side. To address these problems, the survey populates a list of domains (domains that are mapped with any of the user's enabled skills) and asks the user to select all domains that they would have visited.

For the second issue, we define {\em used skills} as the enabled skills the user frequently interacts with. To address potential over-represented skills in the data collected (those enabled but not used), we ask the user to validate the extracted data. The survey also populates all the extracted enabled skills, along with each skill's title and a screenshot of the skill's Amazon homepage, and asks the user to select all skills that they interact with. For both the questions, additional random records ($20$\% of the total records) are added to the populated list. This acts as an attention check for the user. Any user that selects the additional records is removed from the analysis. When the survey detects that it has received the list of apps and skills using the participant's MTurk ID, it guides them to uninstall the extension.

\paragraph{Data Collection Procedure}
We recruited \numUsersTotal individuals through Amazon's MTurk who satisfy the eligibility criteria of age and device ownership. We did not enforce any additional qualifications because the survey has built-in safeguards that ensure the participants correctly installed and executed the extension.  We paid each participant $\$3$ for completing the survey; 95\% of the respondents finished the survey within $9.7$ minutes. The rest of the participants faced some technical issues while installing the extension, and we helped them through the process using email. This communication did not affect the anonymity of the data. Finally, we collected the list of installed Android apps, visited web domains, and preferred skills from these \numUsersTotal individuals. 

We conduct a preliminary check to test whether all the skills from the data collection were included in the Mapper Table. We found that $72$ skills out of $162$ skills were not present as their backlinks were not listed on the websites. We manually verified these skills and added entries to the mapping table. 
% We performed the backlinking through interacting with the official support accounts over email and social media of these skills and disassembling their official mobile apps. 

\subsubsection{Runtime Skill Invocation}
\mjrev{
We perform a runtime evaluation by invoking each skill on a test Amazon Alexa account. This evaluation serves three purposes: (1) perform a large-scale measurement to characterize the phonetic similarity between the skills, (2) confirm Alexa's enable/disable behavior, and (2) assess the security gains for each user in the collected data. We designed an automated and systematic setup to invoke each skill of interest using organic and synthesized speech. We feed an invocation audio command to the Alexa Voice Service to directly interact with the Alexa backend. This invocation command is of the form: ``Alexa, open <skill invocation phrase>.'' We terminate the interaction with the Alexa backend once the skill invocation occurs or if Alexa cannot process the request. We classify the result of each invocation as: ``correct,'' ``incorrect,'' or ``no invocation.'' The last category occurs when Alexa fails to run any skill.
}

\mjrev{We create invocation audio using human speech reconstructed from the LibriSpeech dataset~\cite{libri} and Google TTS (Text-to-Speech service\footnote{\url{cloud.google.com/text-to-speech}}). LibriSpeech consists of transcripts for around 1000 hours of spoken English from $2484$ speakers. To construct an invocation audio from this corpus, we first search for any continuous utterance of the invocation phrase. If nothing is found, we break the invocation phrase into smaller components and stitch together the respective audio clips to create a full phrase. We ensure that the components were derived from the same speaker. Since some skill invocation phrases contain non-English words (e.g., Amex), we were not able to generate organic invocation audio for all skills in the dataset. To avoid any speaker-specific errors, we perform each trial with audio samples collected from three different speakers. Then, we take the majority result from the three outcomes.}

\subsection{Q1. Enable/Disable Module Evaluation}\label{eval_enable_disable}
\sys relies on the Alexa enable/disable API to ensure that the matched skills execute despite the presence of phonetically-close skills. \mjrev{We perform large-scale voice experiments to establish that, when enabling a matched skill and disabling its phonetic neighbors, Alexa always executes the enabled skill. These experiments also inform the distance threshold needed to identify phonetic neighbors.}

\subsubsection{Skill Datasets} The high-level approach is to repeatedly invoke sets of phonetically-close skills and observe the results. We created the following two datasets:\\

\paragraph{Dataset of skills with identical invocation phrases.}\label{identical_eval_skills} We create $8$ disjoint sets of skills where each set contains members with identical invocation phrases. The complete list of skills is in Appendix \ref{appendix}. Note that each skill is randomly taken from the Alexa marketplace (any repetitions in skill names and invocation phrases occur naturally). We have a total of $28$ unique skills. We constructed these sets to test \sys's Enable/Disable module against skills that share invocation phrases with different number of other skills.

    %retrieve $8$ disjoint sets of skills, where each set only contains members with identical invocations. We randomly selected $2$ sets for each identical invocation set of size $2$ through size $5$. Thus, we have $8$ sets and a total of $28$ unique skills in this set of trials. 
    
\paragraph{Dataset of skills with similar sounding invocations.} We select the largest set of skills with unique invocations which satisfies the property --- for each skill, the set also contains at-least two other skills whose invocation phrases are at a phonetic distance of less than $600$. This set of $53$ skills represents a phonetically dense cluster of skills. We select these skills from the whole Alexa market rather than just the set of critical (i.e., account linking) skills to find the ones that are most vulnerable to voice confusion attacks. 

\begin{table}[t]
\caption{The invocation outcomes for skills with \textit{identical} invocation phrases. For the default state, Alexa incorrectly invokes skills. When enabling the target skill and disabling its phonetic neighbors, Alexa has no incorrect invocations. }
\begin{tabular}{p{0.2\columnwidth}cc}
\toprule
 &\multicolumn{2}{c}{\textbf{Text-to-Speech Experiment}}\\
    \cmidrule{2-3}
\quad\quad Metric & Default State & Target Enabled, Others Disabled \\
 \midrule
Correct Invocation     & 2 & 23 \\ 
Incorrect Invocation & 5  & 0  \\
No Invocation          & 1 & 5 \\
\toprule
 &\multicolumn{2}{c}{ \textbf{Audio Recording Experiment}}\\
 \midrule
Correct Invocation     & 2 & 21 \\ 
Incorrect Invocation        & 2  & 0 \\
No Invocation          & 4  & 7 \\
\bottomrule
\end{tabular}\\
% \multicolumn{4}{l}{\makecell{The number of trials varies between TTS and Audio Recording as  \\ not
% all of the invocation phrases are in\\ the LibriSpeech dataset}}
%\multirowcell{6}{The number of trials varies between TTS and Audio Recording as  \\ not
%all of the invocation phrases could be found in\\ the LibriSpeech dataset.}
\label{identical_invocation_experiments}
\end{table}

\begin{table}[t]
\caption{The invocation outcomes for skills with \textit{similar sounding} invocation phrases. For the default state, Alexa incorrectly invokes skills. When enabling the target skill and disabling its phonetic neighbors, Alexa has no incorrect invocations. Note that the number of trials varies between TTS and Audio Recording as not all of the invocation phrases are in the LibriSpeech dataset.}
\begin{tabular}{p{0.2\columnwidth}cc}
\toprule
 &\multicolumn{2}{c}{\textbf{Text-to-Speech Experiment}}\\
    \cmidrule{2-3}
 %\quad\quad Metric & Both Enabled & Target Enabled, Adversary Disabled & Both Disabled \\
 %\midrule
%Correct Invocation     & 87 & 96 & 62 \\ 
%Incorrect Invocation & 5  & 0   & 30  \\
%No Invocation          & 12 & 10  & 14 \\
\quad\quad Metric & Default State & Target Enabled, Adversary Disabled \\
 \midrule
Correct Invocation     & 62 & 96 \\ 
Incorrect Invocation & 30  & 0  \\
No Invocation          & 14 & 10 \\
\toprule
 &\multicolumn{2}{c}{ \textbf{Audio Recording Experiment}}\\
 \midrule
%Correct Invocation     & 40 & 46 & 25 \\ 
%Incorrect Invocation        & 4  & 0  & 13  \\
%No Invocation          & 4  & 2  & 10 \\
Correct Invocation     & 25 & 46 \\ 
Incorrect Invocation        & 13  & 0 \\
No Invocation          & 10  & 2 \\
\bottomrule
\end{tabular}\\
% \multicolumn{4}{l}{\makecell{The number of trials varies between TTS and Audio Recording as  \\ not
% all of the invocation phrases are in\\ the LibriSpeech dataset}}
%\multirowcell{6}{The number of trials varies between TTS and Audio Recording as  \\ not
%all of the invocation phrases could be found in\\ the LibriSpeech dataset.}
\label{tab:exp_table}
\end{table}

\subsubsection{Results for Identical Invocation Phrase Dataset}
We use the Google TTS API and reconstructed human audio to generate audio for invocation phrases, and run these experiments.

%For this set of skills, we use GTTS and AVS to automate interaction with our test Alexa account. The purpose of surveying sets with identical skill invocations and various sizes is to see how Alexa's ASR handles various levels of ambiguity. To this end, we run the following trials for each set: 
\begin{enumerate}
    
    %\item Enable all members of the set, randomly pick one of them as the user's intended invocation and then play audio.
    \item Baseline: With all skills in default state, play audio for invocation phrase. Our intended target is the most popular skill (with most reviews) in the set.
    \item \sys: For each member of the set, enable that member, disable the rest of the set's members, and play audio. The intended target is the enabled skill.
    %\item For N=2 to N=set size, enable N skills and execute invocation.

\end{enumerate}
\mjrev{
For the SkillFence scenario, we report outcomes for $28$ trials while enabling each skill and disabling the rest. However, for the Baseline case, as all the skills are in default state and all skills in one set have identical invocation phrases, we only perform $8$ trials, one for each set. Table~\ref{identical_invocation_experiments} contains the results. We assign the the most popular skill (with most reviews) in the set as the intended target because popularity is a factor in deciding skill invocation~\cite{alexa-no-perm-prompt}. }

\mjrev{
\paragraph{Baseline:} For TTS audio, we observe that in $2$ cases, the user's intended skill is executed (`correct invocation'), in $5$ cases, the unintended skill is invoked and in $1$ case, invocation fails. Similarly, for LibriSpeech audio, there are $2$ cases of 'correct invocation', $2$ cases of 'incorrect invocation' and $4$ cases where invocation fails. Both the experiments conclude that in the presence of identically named skills, confusion does occur.} 

\mjrev{
\paragraph{SkillFence:} For TTS audio, correct invocation of the user's intended skill occurs $23$ times when the intended skill is enabled and unintended ones are disabled ($5$ result in invocation failure). For LibriSpeech audio, there are $22$ correct invocations and in $7$ cases invocation fails. There are no incorrect invocations when the intended skill is enabled and others are disabled. This trial confirms our observed behavior of the enable/disable API --- when the intended skill is enabled and phonetically-close skills are disabled, only the intended skill executes in response to an ambiguous invocation phrase. 
}

\subsubsection{Results for Similar Invocation Phrase Dataset}
In this experiment, we conducted the invocation trials by selecting pairs of skills $(A,B)$, such that skill $B$ is the phonetically closest one to skill $A$. By design, there are $106$ such pairs when we use Google TTS and $48$ such pairs when we use reconstructed human audio. For each pair, we test the invocation of skill $A$ in two configurations:
\begin{enumerate}
    %\item Baseline: Both Enabled (skill $A$ enabled, skill $B$ enabled)
    \item Baseline: Default State (skill $A$ disabled, skill $B$ disabled)
    \item \sys: Target Enabled, Potential Adversary Disabled (skill $A$ enabled, skill $B$ disabled)
    
\end{enumerate} 
Configuration (1) is the control condition that tests how Alexa responds to confusing voice commands currently. Configuration (2) simulates the case where \sys predicts that skill $A$ is the user's intention.  

Table~\ref{tab:exp_table} shows the skill invocation results for both GTTS and organic audio. Configuration (1) incurs $28\%$ and $27\%$ incorrect invocations respectively for the two modes of audio. For configuration (2), the number of incorrect invocations is zero. Thus, we conclude that the behavior of Alexa supports our hypothesis: when a skill is enabled and its phonetic neighbors are disabled, the voice assistant will automatically execute the enabled one. 

%However, for both the sets, Configuration (2) has no incorrect invocations. This demonstrates that \sys  prevents skill squatting even for highly vulnerable skills.  
\begin{figure}[t]
    \centering
    \includegraphics[width=0.5\columnwidth]{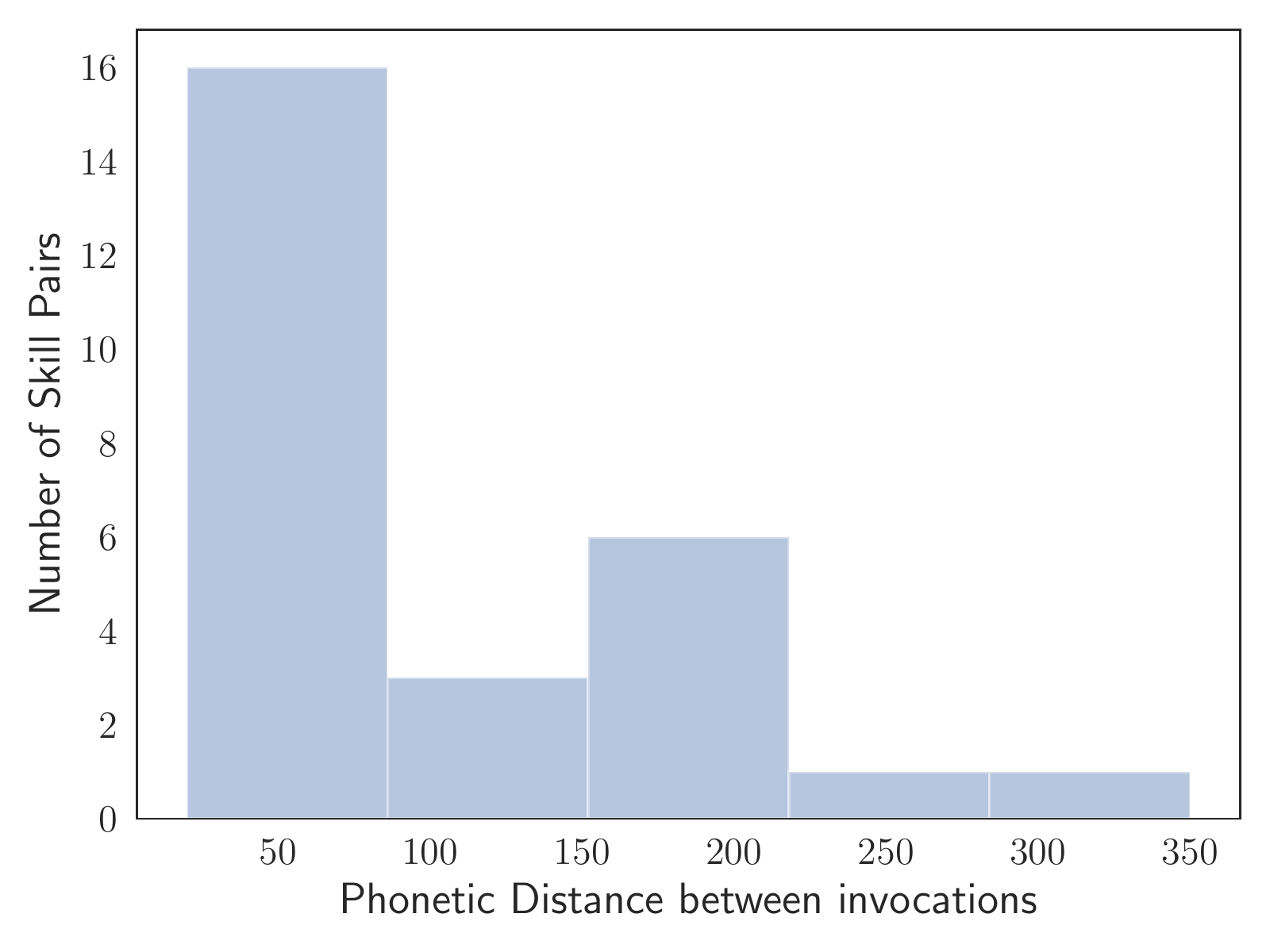}
    \caption{The distribution of phonetic distance between all skill pairs that resulted in incorrect invocation. There are no incorrect pairs beyond a distance of $400$.}
    \label{fig:incorrect_routing_dist}
\end{figure}

\subsubsection{Enable/Disable Threshold}
\label{sec:tuning}
\mjrev{
For the baseline configuration of the similar invocation phrase experiment, there were $30$ and $13$ incorrect invocations for the TTS and human speech trials respectively. Figure~\ref{fig:incorrect_routing_dist} shows the phonetic distance between the invocation of the desired skill and the incorrect skill that was actually invoked. Incorrect invocation occurrences were more frequent between skills whose phonetic distance was smaller. Also, no incorrect invocation occurred between skills with phonetic distance more than $400$. Therefore, for the current set of registered skills on Alexa, \sys, when configured with a threshold greater than $400$ will be able to prevent voice-confusion attacks between any skill pair (assuming the target skill has been mapped). Hence, we fix the phonetic distance threshold for \sys to be $400$ for subsequent experiments.}

\subsection{Q2. End-to-End \sys Performance}
\label{end_to_end_effectiveness}

%define the types of skills. 
\mjrev{We perform a trace-based evaluation of \sys to study its end-to-end performance in terms of the effectiveness in preventing voice confusion attacks, impact on usability, and performance overhead. For each user from our data collection campaign, we generate a list of critical matched skills by running the extension module on the collected counterpart data. Recall that the set of critical skills are ones that are likely to be squatted upon~\cite{shezan-www2020}. This process involves extracting certificates from the visited websites and installed Android apps and then matching the skills from the Mapper table. We create a test Amazon account for each user, enable the matched skills, and disable their phonetic neighbors. We compare the enabled and disabled skills (i.e, those matched from counterpart data and their neighbors) to the skills actually used by the users (reported in the data collection).}

\begin{figure}[t]
    \centering
    \begin{minipage}{.48\textwidth}
        \centering
        \includegraphics[width=0.98\linewidth]{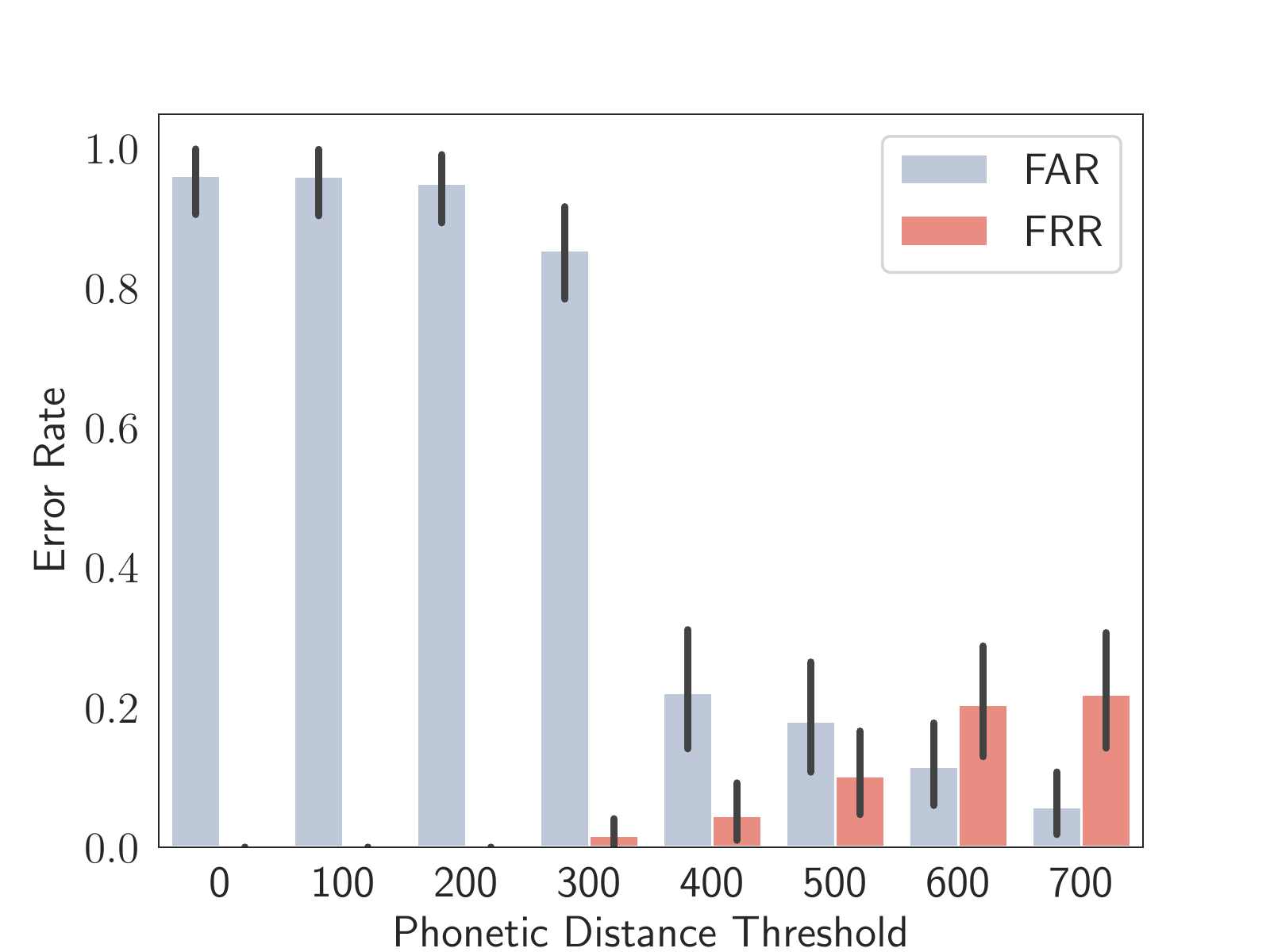}
        \caption{Error Rate and Security-Usability Tradeoff: describes the error rates (FAR and FRR) for different phonetic graph distance thresholds.}
        \label{fig:error_rate}
    \end{minipage}%
    \hfill
    \begin{minipage}{0.48\textwidth}
        \centering
        \includegraphics[width=0.98\linewidth]{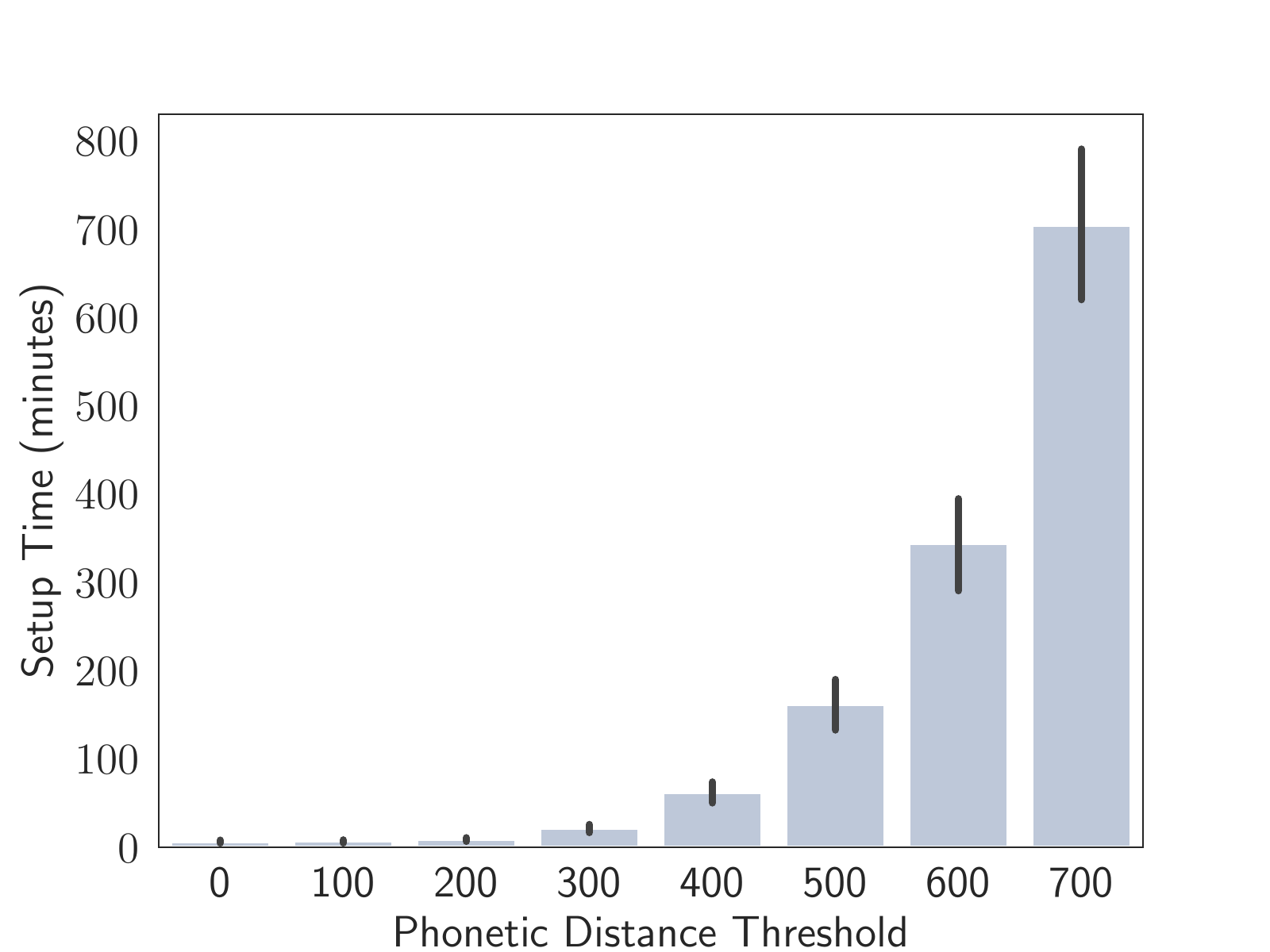}
        \caption{Initial Setup Time: describes \sys's initial setup time for different phonetic graph distance thresholds.}
        \label{fig:setup_time}
    \end{minipage}
\end{figure}

\mjrev{
\subsubsection{Usability vs Security}
 The phonetic distance threshold controls the trade-off between usability and security in \sys. A lower threshold corresponds to a sparser phonetic graph, implying less skills to disable for each matched skill. This setting favors usability as the user can set up \sys faster and can interact with a larger number of skills at the cost of potential incorrect invocations. Whereas, a larger threshold results in a more connected graph with a larger number of skills to be disabled for each matched skill, reducing the invocation confusion and improving security. }
 
 \mjrev{
 We quantify the impact of the phonetic distance threshold through two types of errors: False Rejection Rate (FRR) and False Acceptance Rate (FAR). FRR measures the probability that \sys disables a skill that the user actually needs --- representing a proxy for usability. FAR measures the probability of \sys missing a squatting skill and not disabling it --- representing a proxy for security. Recall that in the previous section we show that a pair of skills with a phonetic distance of less than $400$ can lead to an ambiguous invocation. Using this empirical measure, we consider unused skills, with a distance of less than 400 from any used skill, to be squatting. Figures \ref{fig:error_rate} and \ref{fig:setup_time} show that a lower value of the threshold benefits from a smaller initial setup time and a lower FRR (i.e., better usability). However, a higher threshold improves the security properties by reducing the instances of invocation confusion. We find that a distance threshold of $400$ provides an appropriate trade-off between usability and security. Figure \ref{fig:error_rate} shows that across all users ($N=$ \numUsersTotal), for a distance threshold of $400$, \sys provides an FRR of $9.17$\% and an FAR of \far. Note that the FAR is not $0$\% because \sys did not enable some of the used skills (and correspondingly disable their phonetic neighbors); these skills either did not have a backlink or their counterpart activity was missing.
}

\mjrev{
\subsubsection{Performance Overhead}
Out of all the user-facing modules of \sys, the enabling/disabling phase takes the longest. The primary reason being that we block our extension's operation for two seconds after each new skill page is loaded to avoid Amazon's robot detection that automatically logs the user out. Thus, it takes $2.5$ seconds on average to enable a single skill and $3.08$ seconds to disable a skill. Figure \ref{fig:setup_time} shows that for a distance threshold of $400$, the average initial setup time for a user is about one hour. This operation can be sped up using parallelization at the risk of Amazon blocking \sys. The optimal solution to this problem would be to avoid iteration altogether by using enable and disable functions that allow multiple skills to be selected simultaneously. Should this ability be added, the usability cost for a user's initial defense setup would be greatly reduced. Once the setup is completed, subsequent updates to either the mapper table or the user data only trigger enable/disable of a small number of skills which can be completed in around $2$-$5$ minutes.
}

\begin{figure}[t]
     \centering
     \begin{subfigure}[b]{0.48\textwidth}
         \centering
         \includegraphics[width=\textwidth]{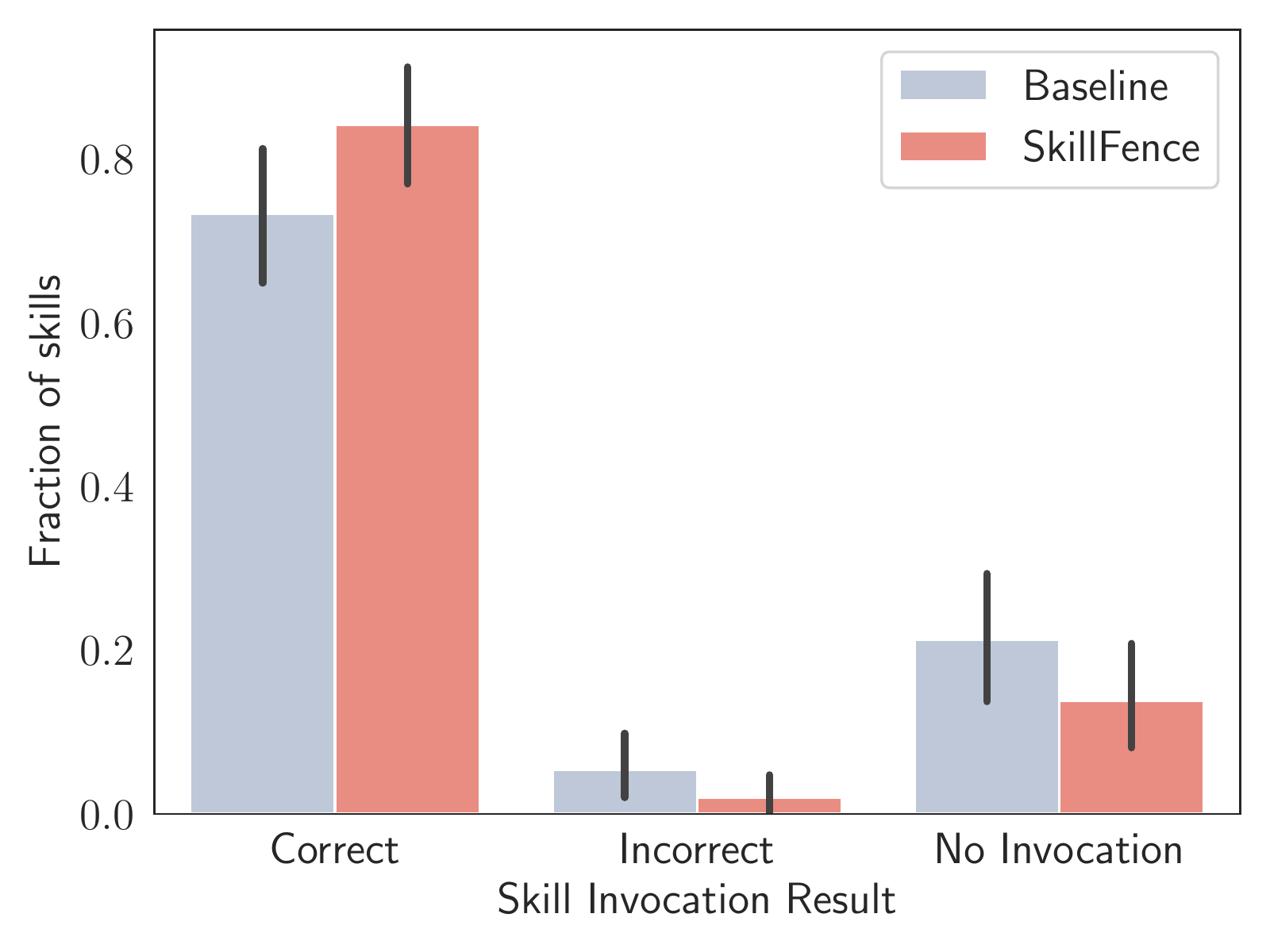}
         \caption{TTS}
         \label{fig:endtendtts}
     \end{subfigure}
     \hfill
     \begin{subfigure}[b]{0.48\textwidth}
         \centering
         \includegraphics[width=\textwidth]{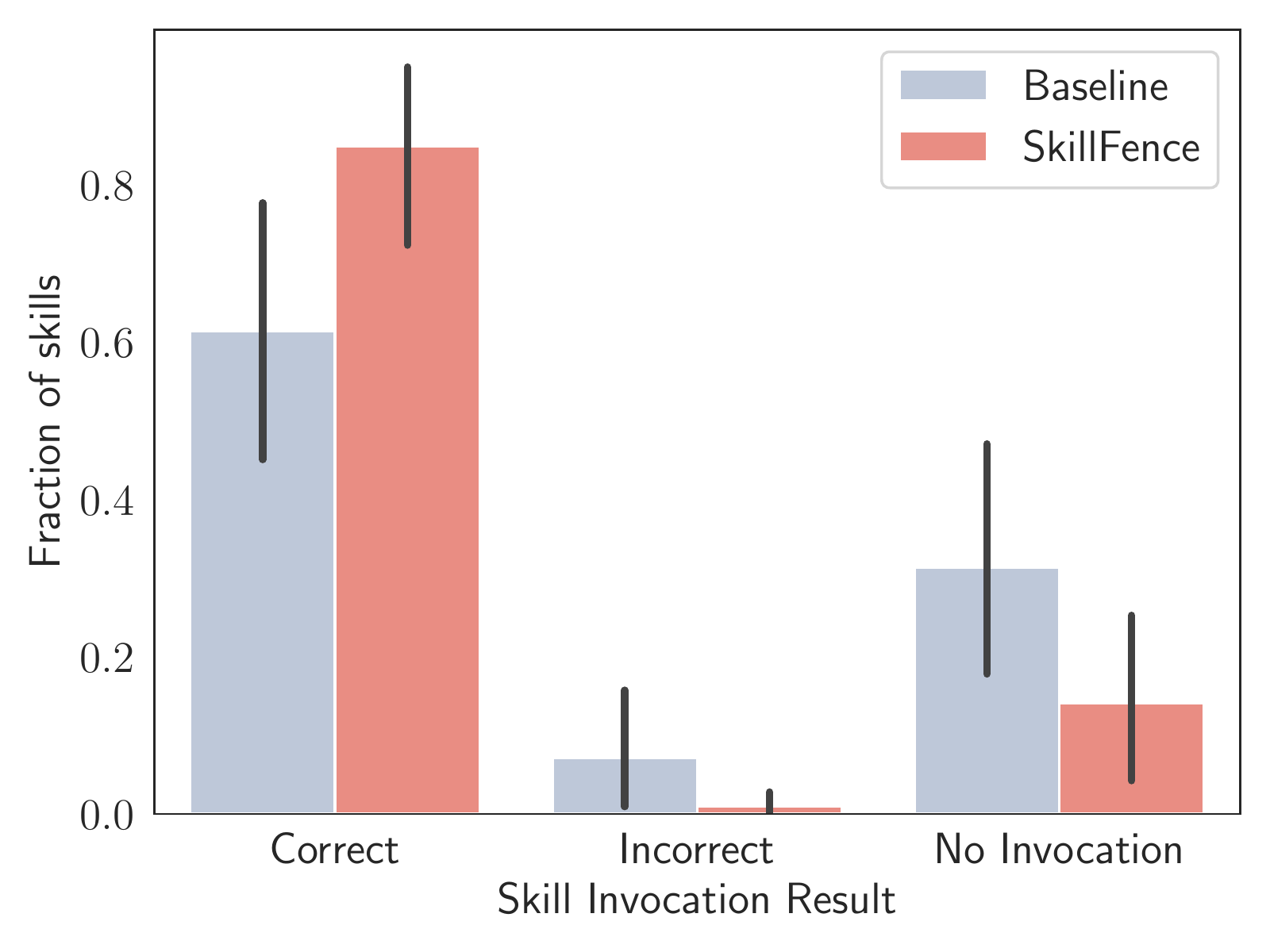}
         \caption{LibriSpeech}
         \label{fig:endtendorg}
     \end{subfigure}
        \caption{Trace based runtime end to end evaluation of \sys. Average Fraction of correct, incorrect and unsuccessful invocations per-user in the user study.}
        \label{fig:endtend}
\end{figure}

%more details here?
\mjrev{
\subsubsection{Runtime Evaluation}
We conclude the trace-based evaluation of \sys by performing invocation trials for all of the users' used skills. For each user, we run these trials in two configurations: baseline and after initializing \sys. To setup \sys, we enable the matched skills and disable their phonetic neighbors within a threshold of $400$. Figure \ref{fig:endtend} shows that \sys is able to increase correct invocations and reduce incorrect invocations for both TTS and human reconstructed (LibriSpeech) audio. On an average, \sys increases correct invocations from $68$\% to $86$\% and reduces incorrect invocations from $6.2$\% to $1.8$\% for a real world user --- the rest corresponds to no invocation instances. Across all users, \sys is able to increase correct invocations from $194$ to $251$, and reduce incorrect invocations from $23$ to $6$. Out of the $6$ incorrect invocations, $4$ were due to the lack of secure backlink and $2$ due to lack of counterpart data. \emph{There are no incorrect invocations for skills with secure backlink and counterpart activity.} Additionally, \sys reduces the instances where Alexa is unable to invoke any skill. 
}
% To set up SkillFence - for each user, enable matched skills, disable their neighbours. (Need to add organic audio results). Then invoke each of the used skills. 
% Without - 89 correct, 10 incorrect, 59 no
% With - 126 correct, 3 incorrect, 29 no

% \ifnum\flagShowFigTable=1
% \begin{figure}[t]
%     \centering
%     \begin{minipage}{.48\textwidth}
%         \centering
%         \includegraphics[width=0.98\linewidth]{images/end-t-end_2.pdf}
%         \caption{TTS}
%         \label{fig:endtendtts}
%     \end{minipage}%
%     \hfill
%     \begin{minipage}{0.48\textwidth}
%         \centering
%         \includegraphics[width=0.98\linewidth]{images/end-t-end_2_organic.pdf}
%         \caption{LibriSpeec}
%         \label{fig:endtendorg}
%     \end{minipage}
%     \caption{Trace based runtime end to end evaluation of \sys. Average Fraction of correct, incorrect and unsuccessful invocations per-user in the user study.}
%     \label{fig:endtend}
% \end{figure}
% \fi

\subsection{Q3. Security and Privacy Analysis}\label{security-analysis}
% reiterate assumptions for security guarantees.
Per our threat model, the attacker is a malicious skill that tries to get executed by leveraging various design flaws in the Alexa ecosystem. The user's browser, smartphone and the legitimate skill developer's websites are trusted and not compromised. An attacker could try to compromise a legitimate website to include a fake link to their attack skill. Because of \sys however, the bar for the attack is now much higher than just uploading a malicious skill to a loosely-vetted marketplace.

% adaptive:
% 0. attacker tries to compromise legit website.
% 1. attacker injects text on website that supports user-generated content to trick backlink analysis
% 2. attacker tricks user into visiting fake website so that their counterpart activity is poisoned
% 3. attacker can try to create a phonetically-close skill JUST outside the phonetic radius threshold we set. 
% 4. privacy-respect collection of web+android history: no transmit

\subsubsection{Adaptive Attack Analysis}
\label{subsubsec:adaptive-attacker}
We consider an adaptive attacker with knowledge of how \sys works. Concretely, this attacker can: (1) trick the user into visiting a fake website such that the counterpart activity  is poisoned; (2) inject text on a legitimate skill developer website that supports user-generated content so that the backlink analysis is tricked; (3) create a skill whose invocation phrase is just outside the phonetic radius of the legitimate skill in an attempt to still leverage voice confusion attacks. As we discuss below, our work prevents these attacks by design. 

\paragraph{Counterpart activity poisoning.} An attacker can trick the user into clicking on websites they control and link those websites to a malicious skill. This can result in malicious skills getting enabled because \sys mistakenly interprets the malicious websites as the user's intention. \sys prevents this attack using its website history filtering technique discussed in Section~\ref{site_history_filter}. An attacker could also try to trick the user into installing an Android app they control. This is a harder task than just uploading a malicious skill to a loosely-vetted marketplace. The attacker has to create an Android app, get it past Google Play's strict vetting policies~\cite{alexavetting, hu2020case} and then trick the user into installing the application. Thus, \sys raises the bar for a voice confusion attack.

\paragraph{Fake URL injection on legitimate website.} If a legitimate skill developer's website supports user-generated content (e.g., a help forum), an attacker can insert a URL to their attack skill, thus tricking \sys into linking the legitimate website's identity with the fake skill. Our backlink analysis prevents this attack by excluding user-generated pages.

\paragraph{Phonetic threshold attack.} The malicious skill developer can choose an invocation phrase that lies just outside the phonetic cluster of an enabled skill. This can lead to a voice-based confusion attack. In Section \ref{eval_enable_disable}, we demonstrate that the probability of incorrect invocation decreases for a higher threshold. Hence, choosing a conservative threshold makes it harder to induce a voice confusion attack.

\subsubsection{Privacy Analysis}
The \sys extension accesses website history and list of installed Android apps. However, it does not store any of this information, not even locally. It also does not transmit it to the server, and only receives the mapper table. It just computes on this information locally using the mapper table. The only information communicated to Amazon is the set of enabled and disabled skills. The set of enabled ones can leak a small amount of information to Amazon if there are skills that user does not eventually use. Concretely, it leaks information that a website corresponding to that skill was visited. This is a small amount of information compared to existing privacy issues with systems like voice assistants~\cite{alexavetting, hu2020case, shezan-www2020} and is reasonable compared to the security advantage of preventing voice confusion attacks.

\color{black}

\section{Design Recommendations}
\label{our_recommendations}

Based on our experience in designing \sys\xspace and evaluating its performance with real user data, we distill the following key recommendations. These recommendations, if implemented by voice assistant manufacturers, broadly limit the impact of voice confusion attacks in the voice assistant ecosystem and improve the efficacy of \sys. We observe that these recommendations are simple to implement by the appropriate stakeholders.

\begin{figure}[t]
 \centerline{
\subfloat[No backlink present on OurGroceries website (before).]
{\includegraphics[width=0.48\columnwidth]{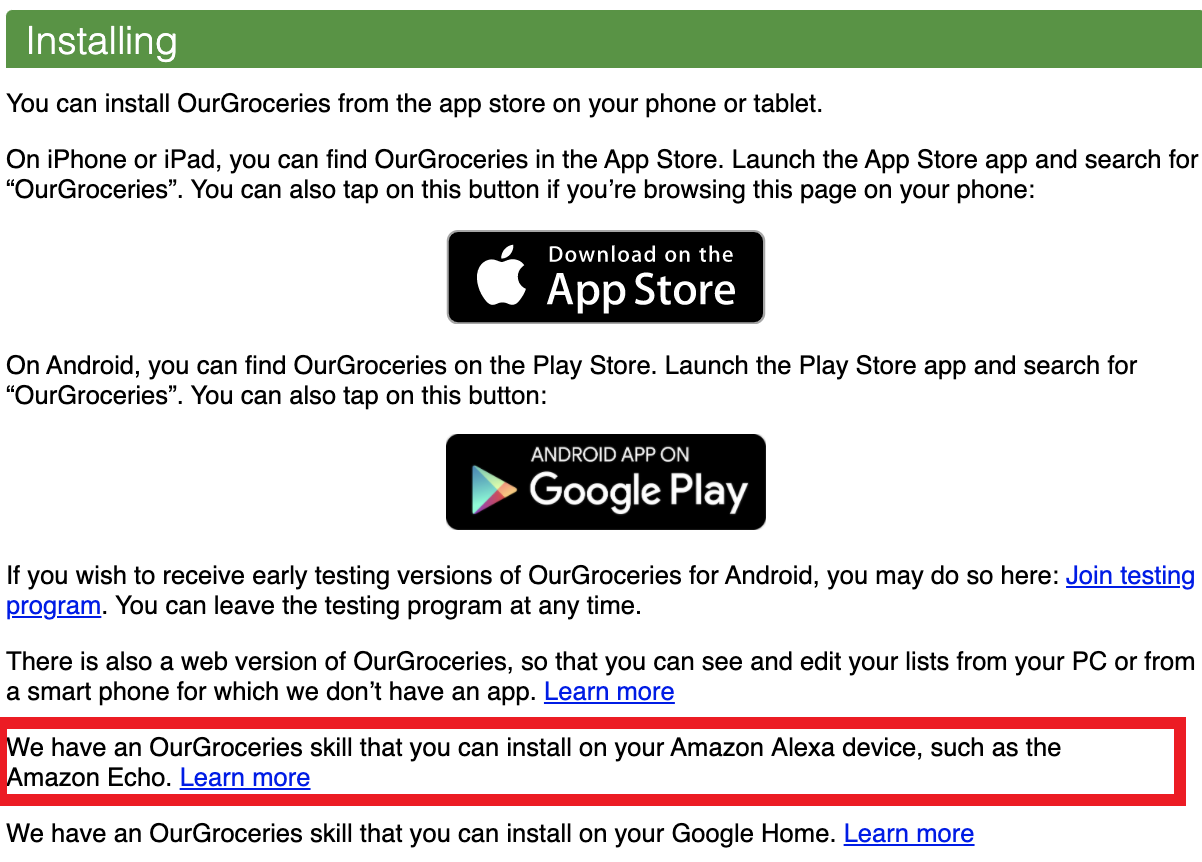}%
%\label{fig:crit_app_correlation}
}
\hfill
\subfloat[OurGroceries added a backlink URL (current).]
{\includegraphics[width=0.48\columnwidth]{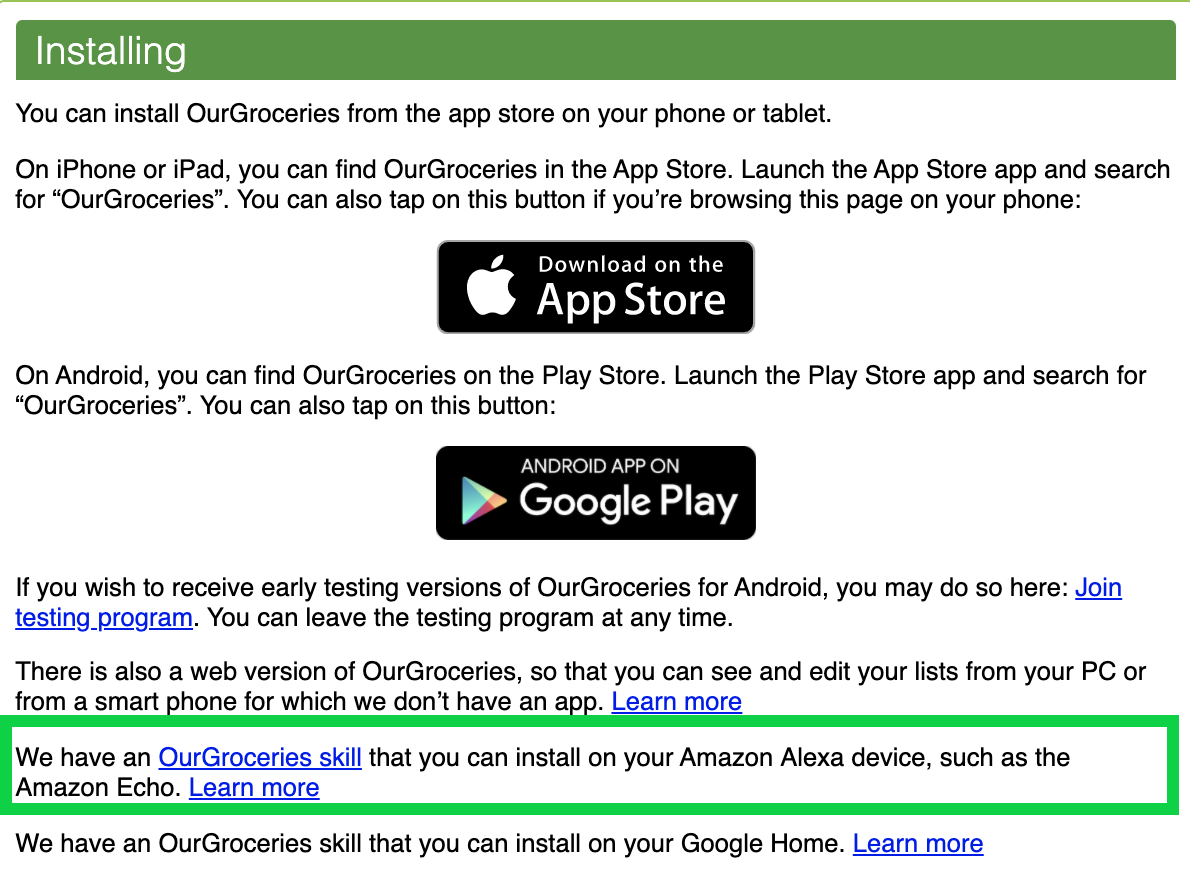}%
%\label{fig:crit_app_correlation_ideal}

}
}
\caption{
% User study - 
% Ratio  of  each user's critical  skills intersecting  with  the  suggested  set  over  the  total  critical  skills for different phonetic distance thresholds}
The updates to \emph{OurGroceries}' website based on our recommendation.
% User Study Graphs
}
\label{fig:our_groceries}
\end{figure} 

\subsection{List Skill Marketplace URLs} We recommend that every skill developer update their website to include a URL to the Amazon marketplace (or the corresponding market in case of other voice assistant manufacturers) listing of their skill. Skill developers already include user guides on their websites for the Alexa skills that they develop. Including this link is a simple change. Indeed, there is precedent for this type of linking --- it is very common for developers to include links to the Android and iOS apps they build for their services. If all developers did this, then \sys\xspace will be able to derive secure identity for all skills in the marketplace. Amazon, as a major stakeholder in the space, can require skill developers to host such a link. A related recommendation is that Amazon also updates their skill marketplace website to include a direct URL to the skill developer's webpage that hosts the backlink. This will remove the need for a backlink search process. 

We began outreach efforts in contacting skill developers to implement this change. We especially credit the \textit{OurGroceries} developers for swiftly implementing this change (Figure \ref{fig:our_groceries}). 

%The before and after for this developer's website\footnote{\url{https://www.ourgroceries.com/user-guide\#installing}} is shown in .

\subsection{Validate Skill Metadata} We recommend that voice assistant manufacturers like Alexa properly vet the metadata of skills on the marketplace. For example, if a skill developer lists a certain URL for their privacy policy, Amazon should perform domain-ownership testing to ensure that indeed, this skill developer owns the domain where the privacy policy is hosted. This test will limit the ability of attackers to mirror legitimate skills. However, this recommendation represents a larger infrastructural and procedural change to the Alexa ecosystem compared to the first recommendation above.

\subsection{Provide Better Skill Invocation Control} We recommend that voice assistant platforms provide an efficient API to control skill invocation.  Systems like \sys\xspace will benefit from an API that can batch-allow a certain set of skills to be eligible for execution while batch-disabling a set of skills to be \emph{ineligible} for execution despite confusing voice commands. Our work extracts this property from Alexa's skill enable/disable API, but its inflexible nature requires optimizations that trade-off performance and security. We also recommend that Amazon clarify the behavior of their Enable and Disable mechanisms. In this work, we empirically verify that the Enable primitive prioritizes the enabled skill over other skills. However, a concrete understanding of both the extent to which an enabled skill is preferred and the  effects of enabling multiple skills (phonetically similar or not) cannot be gained without an explanation of how the enable primitive interacts with Alexa's ASR. 
%Similarly, we show throughout our experiments that the disable mechanism does not allow a skill to be executed. But, we cannot guarantee this behavior in all possible circumstances until the technical details of the disable mechanism's construction are released. 
Furthermore, if the details of the enable/disable mechanisms are published, \sys\xspace will be able to dispense security recommendations that are globally applicable and deal with every conceivable interaction with these two primitives.

\section{Limitations}
\mjrev{
\subsection{Applicability to Other \avP} 
 Our design and implementation of \sys consider Alexa because it is the most popular voice assistant system. 
 It requires three components: an invocation phrase, a notion of secure identity, and control over which skill executes. These components apply to the Google Voice Assistant, the other popular voice assistant system. 
 First, the Google Voice Assistant, allows users to interact with apps (called actions) using their names - ``Let me speak/talk to $<$app name$>$''. The conflicts from the action name induce uncertainties for this platform as well. Second, the Google voice assistant ecosystem lists the actions and links to developer websites and privacy policies. We verified that a similar notion of secure identity
 could be achieved using the search and crawl approach.\footnote{Starbucks is an example: 
 \url{www.starbucks.com/promo/googleassistant}} Third, the  Google Assistant has a subset of actions that allow users to link their accounts. The users can also unlink a linked action. It is, however, unclear how the link/unlink operation affects the action invocation process --- this will require additional experimentation that we leave to future work.}
 
\mjrev{
\subsection{Low Availability of Secure Backlinks} We were able to find secure backlinks for $474$ out of $3659$ skills on the Amazon Alexa market. Although the proportion ($72$ out of $162$ skills) was higher for skills actually used by real world users (from the user study), it is still inadequate. However, this number is likely to increase as almost all developers list android/iOS app links on their websites --- it is only a matter of time before they start adding links for their Alexa skills. 
We are also engaging in outreach efforts by contacting developers and helping them understand the value of adding backlinks to their websites. Upon our recommendation, the developers of OurGroceries (Figure ~\ref{fig:our_groceries}) have already updated their website to include a backlink to their Alexa skill.
}

\mjrev{
\subsection{Usability Study} 
% what would the usability study comprise of? 
% Why is it out of scope?
% What did we do instead?
One limitation of our evaluation is that we do not directly evaluate users' experiences with \sys. Such an analysis would involve tracking user interactions with \sys and the Alexa skill ecosystem over an extended period of time. While it would provide useful insight into \sys's adoption and continued usage rates, such study is out of scope for this work; in this paper, we lay out \sys's motivation, design and effectiveness. Hence, we leave that to future work. However, we do perform an  evaluation of \sys's usability by relying on standard metrics like the False Acceptance Rate (FAR) and False Rejection Rate (FRR). We also study the overhead incurred by the end-users while using \sys in terms of the initial setup time. 
}

\mjrev{
\subsection{Enable/Disable Setup Time} \sys requires an average initial setup time of about one hour. The primary reason for this overhead is that the Alexa skill API only offers a method to enable or disable a single skill at a time. Attempts to parallelize these operations run the risk of Amazon blocking \sys. \sys tries to address this usability overhead by performing most of the operations in the background. Also, once the setup is completed, later updates can be perform much faster (around $2$-$5$ minutes). We recommend that voice assistant manufacturers provide an API that can efficiently disable a large list of skills with a single network call.
}
\mjrev{
\section{Conclusion}
We propose \sys, a browser extension that prevents voice-based confusion attacks on voice assistants like Amazon Alexa. Our system analyzes a user's counterpart activity to reduce dis-ambiguities in the skill invocation process and ensure only the skills preferred by the user execute despite ambiguity. Using real user data, we evaluate the different components of \sys\xspace and demonstrate its utility in protecting existing users. We distill recommendations for stakeholders that, if adopted, will increase the effectiveness of \sys, and we are beginning to see adoption.
}

%We also propose recommendations to improve the overall security of the Alexa ecosystem, while also increasing the effectiveness of \sys\xspace.

%characterizes skill usage by predicting skills based on users' counterpart activity on other devices such as the web and smartphones, for which it then ensures correct invocation even under ambiguous voice commands. 

\begin{acks}

We thank the anonymous reviewers for their constructive feedback that has made the work stronger. This work was supported in part by the University of Wisconsin-Madison Office of the Vice Chancellor for Research and Graduate Education with funding from the Wisconsin Alumni Research Foundation. This work was also supported in part by DARPA (through the GARD program) and the NSF through awards: CNS-1838733, CNS-1942014, and CNS-2003129.
\end{acks}

%\begin{acks}
%To Robert, for the bagels and explaining CMYK and color spaces.
%\end{acks}

\bibliographystyle{ACM-Reference-Format}
\bibliography{ref_new}


\begin{thebibliography}{29}

%%% the Web address.

\ifx \showCODEN    \undefined \def \showCODEN     #1{\unskip}     \fi
\ifx \showDOI      \undefined \def \showDOI       #1{#1}\fi
\ifx \showISBNx    \undefined \def \showISBNx     #1{\unskip}     \fi
\ifx \showISBNxiii \undefined \def \showISBNxiii  #1{\unskip}     \fi
\ifx \showISSN     \undefined \def \showISSN      #1{\unskip}     \fi
\ifx \showLCCN     \undefined \def \showLCCN      #1{\unskip}     \fi
\ifx \shownote     \undefined \def \shownote      #1{#1}          \fi
\ifx \showarticletitle \undefined \def \showarticletitle #1{#1}   \fi
\ifx \showURL      \undefined \def \showURL       {\relax}        \fi
% invisible to TeX
\providecommand\bibfield[2]{#2}
\providecommand\bibinfo[2]{#2}
\providecommand\natexlab[1]{#1}
\providecommand\showeprint[2][]{arXiv:#2}

\bibitem[Carlini et~al\mbox{.}(2016)]%
        {hvc}
\bibfield{author}{\bibinfo{person}{N. Carlini}, \bibinfo{person}{P. Mishra},
  \bibinfo{person}{T. Vaidya}, \bibinfo{person}{Y. Zhang}, \bibinfo{person}{M.
  Sherr}, \bibinfo{person}{C. Shields}, \bibinfo{person}{D. Wagner}, {and}
  \bibinfo{person}{W. Zhou}.} \bibinfo{year}{2016}\natexlab{}.
\newblock \showarticletitle{Hidden Voice Commands}. In
  \bibinfo{booktitle}{\emph{25th USENIX Security Symposium (USENIX Security
  16)}}.
\newblock


\bibitem[Chen et~al\mbox{.}(2020)]%
        {devil}
\bibfield{author}{\bibinfo{person}{Yuxuan Chen}, \bibinfo{person}{Xuejing
  Yuan}, \bibinfo{person}{Jiangshan Zhang}, \bibinfo{person}{Yue Zhao},
  \bibinfo{person}{Kai Zhang, Shengzhi~Chen}, {and} \bibinfo{person}{XiaoFeng
  Wang}.} \bibinfo{year}{2020}\natexlab{}.
\newblock \showarticletitle{Devil{\textquoteright}s Whisper: A General Approach
  for Physical Adversarial Attacks against Commercial Black-box Speech
  Recognition Devices}. In \bibinfo{booktitle}{\emph{29th {USENIX} Security
  Symposium ({USENIX} Security 20)}}.
\newblock


\bibitem[{E. Fernandes} et~al\mbox{.}(2016)]%
        {flowfence16}
\bibfield{author}{\bibinfo{person}{{E. Fernandes}}, \bibinfo{person}{J.
  Paupore}, \bibinfo{person}{A. Rahmati}, \bibinfo{person}{D. Simionato},
  \bibinfo{person}{M. Conti}, {and} \bibinfo{person}{A. Prakash}.}
  \bibinfo{year}{2016}\natexlab{}.
\newblock \showarticletitle{{F}low{F}ence: {P}ractical {D}ata {P}rotection for
  {E}merging {I}o{T} {A}pplication {F}rameworks}. In
  \bibinfo{booktitle}{\emph{Proceedings of the 25th {USENIX} Security
  Symposium}}.
\newblock


\bibitem[Gerasimenko(2020)]%
        {Gerasimenko:2010}
\bibfield{author}{\bibinfo{person}{Dmitry Gerasimenko}.} \bibinfo{year}{2010
  (accessed 2020)}\natexlab{}.
\newblock \bibinfo{booktitle}{\emph{Ahrefs}}.
\newblock
\urldef\tempurl%
\url{https://ahrefs.com}
\showURL{%
\tempurl}


\bibitem[Guo et~al\mbox{.}(2020)]%
        {skillex}
\bibfield{author}{\bibinfo{person}{Zhixiu Guo}, \bibinfo{person}{Zijin Lin},
  \bibinfo{person}{Pan Li}, {and} \bibinfo{person}{Kai Chen}.}
  \bibinfo{year}{2020}\natexlab{}.
\newblock \showarticletitle{SkillExplorer: Understanding the Behavior of Skills
  in Large Scale}. In \bibinfo{booktitle}{\emph{29th {USENIX} Security
  Symposium ({USENIX} Security 20)}}. \bibinfo{publisher}{{USENIX}
  Association}.
\newblock
\urldef\tempurl%
\url{https://www.usenix.org/conference/usenixsecurity20/presentation/guo}
\showURL{%
\tempurl}


\bibitem[Hu et~al\mbox{.}(2020)]%
        {hu2020case}
\bibfield{author}{\bibinfo{person}{Hang Hu}, \bibinfo{person}{Limin Yang},
  \bibinfo{person}{Shihan Lin}, {and} \bibinfo{person}{Gang Wang}.}
  \bibinfo{year}{2020}\natexlab{}.
\newblock \showarticletitle{A Case Study of the Security Vetting Process of
  Smart-home Assistant Applications}. In \bibinfo{booktitle}{\emph{IEEE
  Workshop on the Internet of Safe Things (SafeThings)}}.
\newblock


\bibitem[Inc.({[n.\,d.]})]%
        {alexa-guidelines}
\bibfield{author}{\bibinfo{person}{Amazon Inc.}}
  \bibinfo{year}{[n.\,d.]}\natexlab{}.
\newblock \bibinfo{title}{{Alexa Skill Certification Guidelines}}.
\newblock
  \bibinfo{howpublished}{\url{https://developer.amazon.com/en-US/docs/alexa/custom-skills/certification-requirements-for-custom-skills.html}}.
\newblock


\bibitem[Kinsella(2018)]%
        {voicebot}
\bibfield{author}{\bibinfo{person}{BRET Kinsella}.}
  \bibinfo{year}{2018}\natexlab{}.
\newblock \bibinfo{title}{{Should Amazon Alexa Stop Allowing Duplicate
  Invocation Names? Should Google Assistant Permit Them?}}
\newblock
  \bibinfo{howpublished}{\url{https://voicebot.ai/2018/03/26/amazon-alexa-stop-allowing-duplicate-invocation-names-google-assistant-permit/}}.
\newblock


\bibitem[Kumar et~al\mbox{.}(2018a)]%
        {kumar}
\bibfield{author}{\bibinfo{person}{Deepak Kumar}, \bibinfo{person}{Riccardo
  Paccagnella}, \bibinfo{person}{Paul Murley}, \bibinfo{person}{Eric
  Hennenfent}, \bibinfo{person}{Joshua Mason}, \bibinfo{person}{Adam Bates},
  {and} \bibinfo{person}{Michael Bailey}.} \bibinfo{year}{2018}\natexlab{a}.
\newblock \showarticletitle{Skill Squatting Attacks on Amazon Alexa}. In
  \bibinfo{booktitle}{\emph{27th {USENIX} Security Symposium ({USENIX} Security
  18)}}. \bibinfo{publisher}{{USENIX} Association},
  \bibinfo{address}{Baltimore, MD}, \bibinfo{pages}{33--47}.
\newblock
\showISBNx{978-1-931971-46-1}
\urldef\tempurl%
\url{https://www.usenix.org/conference/usenixsecurity18/presentation/kumar}
\showURL{%
\tempurl}


\bibitem[Kumar et~al\mbox{.}(2018b)]%
        {kumar2018skill}
\bibfield{author}{\bibinfo{person}{Deepak Kumar}, \bibinfo{person}{Riccardo
  Paccagnella}, \bibinfo{person}{Paul Murley}, \bibinfo{person}{Eric
  Hennenfent}, \bibinfo{person}{Joshua Mason}, \bibinfo{person}{Adam Bates},
  {and} \bibinfo{person}{Michael Bailey}.} \bibinfo{year}{2018}\natexlab{b}.
\newblock \showarticletitle{Skill squatting attacks on Amazon Alexa}. In
  \bibinfo{booktitle}{\emph{27th $\{$USENIX$\}$ Security Symposium
  ($\{$USENIX$\}$ Security 18)}}. \bibinfo{pages}{33--47}.
\newblock


\bibitem[Lentzsch et~al\mbox{.}(2021a)]%
        {alexavetting}
\bibfield{author}{\bibinfo{person}{Christopher Lentzsch},
  \bibinfo{person}{Sheel~Jayesh Shah}, \bibinfo{person}{Benjamin Andow},
  \bibinfo{person}{Martin Degeling}, \bibinfo{person}{Anupam Das}, {and}
  \bibinfo{person}{William Enck}.} \bibinfo{year}{2021}\natexlab{a}.
\newblock \showarticletitle{Hey {Alexa}, is this Skill Safe?: Taking a Closer
  Look at the {Alexa} Skill Ecosystem}. In
  \bibinfo{booktitle}{\emph{Proceedings of the 28th ISOC Annual Network and
  Distributed Systems Symposium (NDSS)}}.
\newblock


\bibitem[Lentzsch et~al\mbox{.}(2021b)]%
        {alexa-skill-ecosystem-2021}
\bibfield{author}{\bibinfo{person}{Christopher Lentzsch},
  \bibinfo{person}{Sheel~Jayesh Shah}, \bibinfo{person}{Benjamin Andow},
  \bibinfo{person}{Martin Degeling}, \bibinfo{person}{Anupam Das}, {and}
  \bibinfo{person}{William Enck}.} \bibinfo{year}{2021}\natexlab{b}.
\newblock \showarticletitle{Hey {Alexa}, is this Skill Safe?: Taking a Closer
  Look at the {Alexa} Skill Ecosystem}. In
  \bibinfo{booktitle}{\emph{Proceedings of the 28th ISOC Annual Network and
  Distributed Systems Symposium (NDSS)}}.
\newblock


\bibitem[Liu et~al\mbox{.}(2008)]%
        {history-filter}
\bibfield{author}{\bibinfo{person}{Yiqun Liu}, \bibinfo{person}{Rongwei Cen},
  \bibinfo{person}{Min Zhang}, \bibinfo{person}{Shaoping Ma}, {and}
  \bibinfo{person}{Liyun Ru}.} \bibinfo{year}{2008}\natexlab{}.
\newblock \showarticletitle{Identifying Web Spam with User Behavior Analysis}.
  In \bibinfo{booktitle}{\emph{Proceedings of the 4th International Workshop on
  Adversarial Information Retrieval on the Web}} (Beijing, China)
  \emph{(\bibinfo{series}{AIRWeb '08})}. \bibinfo{publisher}{Association for
  Computing Machinery}, \bibinfo{address}{New York, NY, USA},
  \bibinfo{pages}{9–16}.
\newblock
\showISBNx{9781605581590}
\urldef\tempurl%
\url{https://doi.org/10.1145/1451983.1451986}
\showDOI{\tempurl}


\bibitem[Major et~al\mbox{.}(2019)]%
        {major2019alexa}
\bibfield{author}{\bibinfo{person}{David~J. Major},
  \bibinfo{person}{Danny~Yuxing Huang}, \bibinfo{person}{Marshini Chetty},
  {and} \bibinfo{person}{Nick Feamster}.} \bibinfo{year}{2019}\natexlab{}.
\newblock \bibinfo{title}{Alexa, Who Am I Speaking To? Understanding Users'
  Ability to Identify Third-Party Apps on Amazon Alexa}.
\newblock
\newblock
\showeprint[arxiv]{1910.14112}~[cs.HC]


\bibitem[{Panayotov} et~al\mbox{.}(2015)]%
        {libri}
\bibfield{author}{\bibinfo{person}{V. {Panayotov}}, \bibinfo{person}{G.
  {Chen}}, \bibinfo{person}{D. {Povey}}, {and} \bibinfo{person}{S.
  {Khudanpur}}.} \bibinfo{year}{2015}\natexlab{}.
\newblock \showarticletitle{Librispeech: An ASR corpus based on public domain
  audio books}. In \bibinfo{booktitle}{\emph{2015 IEEE International Conference
  on Acoustics, Speech and Signal Processing (ICASSP)}}.
  \bibinfo{pages}{5206--5210}.
\newblock
\urldef\tempurl%
\url{https://doi.org/10.1109/ICASSP.2015.7178964}
\showDOI{\tempurl}


\bibitem[{Paul Cutsinger}(2018)]%
        {alexa-no-perm-prompt}
\bibfield{author}{\bibinfo{person}{{Paul Cutsinger}}.}
  \bibinfo{year}{2018}\natexlab{}.
\newblock \bibinfo{title}{How to Improve Alexa Skill Discovery with Name-Free
  Interaction and More}.
\newblock
  \bibinfo{howpublished}{\url{https://developer.amazon.com/blogs/alexa/post/0fecdb38-97c9-48ac-953b-23814a469cfc/skill-discovery}}.
\newblock


\bibitem[{Ritik Singh}(2021)]%
        {android_backlink}
\bibfield{author}{\bibinfo{person}{{Ritik Singh}}.}
  \bibinfo{year}{2021}\natexlab{}.
\newblock \bibinfo{title}{7 Ways to Find If an App Is Fake or Real Before
  Installing It}.
\newblock
  \bibinfo{howpublished}{\url{https://gadgetstouse.com/blog/2021/04/19/find-app-is-fake-or-real-before-installing/}}.
\newblock


\bibitem[Roy et~al\mbox{.}(2018)]%
        {inaudble}
\bibfield{author}{\bibinfo{person}{Nirupam Roy}, \bibinfo{person}{Sheng Shen},
  \bibinfo{person}{Haitham Hassanieh}, {and} \bibinfo{person}{Romit~Roy
  Choudhury}.} \bibinfo{year}{2018}\natexlab{}.
\newblock \showarticletitle{Inaudible Voice Commands: The Long-Range Attack and
  Defense}. In \bibinfo{booktitle}{\emph{15th {USENIX} Symposium on Networked
  Systems Design and Implementation ({NSDI} 18)}}. \bibinfo{publisher}{{USENIX}
  Association}, \bibinfo{address}{Renton, WA}, \bibinfo{pages}{547--560}.
\newblock
\showISBNx{978-1-939133-01-4}
\urldef\tempurl%
\url{https://www.usenix.org/conference/nsdi18/presentation/roy}
\showURL{%
\tempurl}


\bibitem[Shezan et~al\mbox{.}(2020)]%
        {shezan-www2020}
\bibfield{author}{\bibinfo{person}{Faysal~Hossain Shezan},
  \bibinfo{person}{Hang Hu}, \bibinfo{person}{Jiamin Wang},
  \bibinfo{person}{Gang Wang}, {and} \bibinfo{person}{Yuan Tian}.}
  \bibinfo{year}{2020}\natexlab{}.
\newblock \showarticletitle{Read Between the Lines: An Empirical Measurement of
  Sensitive Applications of Voice Personal Assistant Systems}. In
  \bibinfo{booktitle}{\emph{Proceedings of the Web Conference (WWW'20)}}.
\newblock


\bibitem[Sugawara et~al\mbox{.}(2020)]%
        {sugawara2020light}
\bibfield{author}{\bibinfo{person}{Takeshi Sugawara}, \bibinfo{person}{Benjamin
  Cyr}, \bibinfo{person}{Sara Rampazzi}, \bibinfo{person}{Daniel Genkin}, {and}
  \bibinfo{person}{Kevin Fu}.} \bibinfo{year}{2020}\natexlab{}.
\newblock \showarticletitle{Light Commands: Laser-Based Audio Injection Attacks
  on Voice-Controllable Systems}. In \bibinfo{booktitle}{\emph{29th {USENIX}
  Security Symposium ({USENIX} Security 20)}}.
\newblock


\bibitem[{Suresh} and {Padmajavalli}(2007)]%
        {4221889}
\bibfield{author}{\bibinfo{person}{R.~M. {Suresh}} {and} \bibinfo{person}{R.
  {Padmajavalli}}.} \bibinfo{year}{2007}\natexlab{}.
\newblock \showarticletitle{An Overview of Data Preprocessing in Data and Web
  Usage Mining}. In \bibinfo{booktitle}{\emph{2006 1st International Conference
  on Digital Information Management}}. \bibinfo{pages}{193--198}.
\newblock
\urldef\tempurl%
\url{https://doi.org/10.1109/ICDIM.2007.369352}
\showDOI{\tempurl}


\bibitem[Vaidya et~al\mbox{.}(2015)]%
        {noodles}
\bibfield{author}{\bibinfo{person}{Tavish Vaidya}, \bibinfo{person}{Yuankai
  Zhang}, \bibinfo{person}{Micah Sherr}, {and} \bibinfo{person}{Clay Shields}.}
  \bibinfo{year}{2015}\natexlab{}.
\newblock \showarticletitle{Cocaine noodles: exploiting the gap between human
  and machine speech recognition}. In \bibinfo{booktitle}{\emph{9th
  $\{$USENIX$\}$ Workshop on Offensive Technologies ($\{$WOOT$\}$ 15)}}.
\newblock


\bibitem[{voicebot.ai}(2021)]%
        {skill_stats}
\bibfield{author}{\bibinfo{person}{{voicebot.ai}}.}
  \bibinfo{year}{2021}\natexlab{}.
\newblock \bibinfo{title}{Alexa Skill Counts Surpass 80K in US, Spain Adds the
  Most Skills, New Skill Rate Falls Globally}.
\newblock
  \bibinfo{howpublished}{\url{https://voicebot.ai/2021/01/14/alexa-skill-counts-surpass-80k-in-us-spain-adds-the-most-skills-new-skill-introduction-rate-continues-to-fall-across-countries/}}.
\newblock


\bibitem[Yuan et~al\mbox{.}(2018)]%
        {commander-song}
\bibfield{author}{\bibinfo{person}{Xuejing Yuan}, \bibinfo{person}{Yuxuan
  Chen}, \bibinfo{person}{Yue Zhao}, \bibinfo{person}{Yunhui Long},
  \bibinfo{person}{Xiaokang Liu}, \bibinfo{person}{Kai Chen},
  \bibinfo{person}{Shengzhi Zhang}, \bibinfo{person}{Heqing Huang},
  \bibinfo{person}{XiaoFeng Wang}, {and} \bibinfo{person}{Carl~A. Gunter}.}
  \bibinfo{year}{2018}\natexlab{}.
\newblock \showarticletitle{Commandersong: A Systematic Approach for Practical
  Adversarial Voice Recognition}. In \bibinfo{booktitle}{\emph{Proceedings of
  the 27th USENIX Conference on Security Symposium}} (Baltimore, MD, USA)
  \emph{(\bibinfo{series}{SEC’18})}. \bibinfo{publisher}{USENIX Association},
  \bibinfo{address}{USA}, \bibinfo{pages}{49–64}.
\newblock
\showISBNx{9781931971461}


\bibitem[Zhang et~al\mbox{.}(2017)]%
        {dolphin}
\bibfield{author}{\bibinfo{person}{Guoming Zhang}, \bibinfo{person}{Chen Yan},
  \bibinfo{person}{Xiaoyu Ji}, \bibinfo{person}{Tianchen Zhang},
  \bibinfo{person}{Taimin Zhang}, {and} \bibinfo{person}{Wenyuan Xu}.}
  \bibinfo{year}{2017}\natexlab{}.
\newblock \showarticletitle{DolphinAttack: Inaudible Voice Commands}. In
  \bibinfo{booktitle}{\emph{Proceedings of the 2017 ACM SIGSAC Conference on
  Computer and Communications Security}} (Dallas, Texas, USA)
  \emph{(\bibinfo{series}{CCS ’17})}. \bibinfo{publisher}{Association for
  Computing Machinery}, \bibinfo{address}{New York, NY, USA},
  \bibinfo{pages}{103–117}.
\newblock
\showISBNx{9781450349468}
\urldef\tempurl%
\url{https://doi.org/10.1145/3133956.3134052}
\showDOI{\tempurl}


\bibitem[Zhang et~al\mbox{.}(2019a)]%
        {zhang}
\bibfield{author}{\bibinfo{person}{N. Zhang}, \bibinfo{person}{X. Mi},
  \bibinfo{person}{X. Feng}, \bibinfo{person}{X. Wang}, \bibinfo{person}{Y.
  Tian}, {and} \bibinfo{person}{F. Qian}.} \bibinfo{year}{2019}\natexlab{a}.
\newblock \showarticletitle{Dangerous Skills: Understanding and Mitigating
  Security Risks of Voice-Controlled Third-Party Functions on Virtual Personal
  Assistant Systems}. In \bibinfo{booktitle}{\emph{2019 IEEE Symposium on
  Security and Privacy (SP)}}, Vol.~\bibinfo{volume}{00}.
  \bibinfo{pages}{263--278}.
\newblock
\showISSN{CFP19020-ART}
\urldef\tempurl%
\url{https://doi.org/10.1109/SP.2019.00016}
\showDOI{\tempurl}


\bibitem[Zhang et~al\mbox{.}(2019b)]%
        {lipfuzzer}
\bibfield{author}{\bibinfo{person}{Yangyong Zhang}, \bibinfo{person}{Lei Xu},
  \bibinfo{person}{Abner Mendoza}, \bibinfo{person}{Guangliang Yang},
  \bibinfo{person}{Phakpoom Chinprutthiwong}, {and} \bibinfo{person}{Guofei
  Gu}.} \bibinfo{year}{2019}\natexlab{b}.
\newblock \showarticletitle{Life after Speech Recognition: Fuzzing Semantic
  Misinterpretation for Voice Assistant Applications}. In
  \bibinfo{booktitle}{\emph{Proceedings of the Network and Distributed System
  Security Symposium (NDSS'19)}}.
\newblock


\bibitem[Zhang et~al\mbox{.}(2019c)]%
        {zhang2019life}
\bibfield{author}{\bibinfo{person}{Yangyong Zhang}, \bibinfo{person}{Lei Xu},
  \bibinfo{person}{Abner Mendoza}, \bibinfo{person}{Guangliang Yang},
  \bibinfo{person}{Phakpoom Chinprutthiwong}, {and} \bibinfo{person}{Guofei
  Gu}.} \bibinfo{year}{2019}\natexlab{c}.
\newblock \showarticletitle{Life after Speech Recognition: Fuzzing Semantic
  Misinterpretation for Voice Assistant Applications.}. In
  \bibinfo{booktitle}{\emph{NDSS}}.
\newblock


\bibitem[{Zhou} et~al\mbox{.}(2006)]%
        {4061378}
\bibfield{author}{\bibinfo{person}{B. {Zhou}}, \bibinfo{person}{S.~C. {Hui}},
  {and} \bibinfo{person}{A.~C. m. {Fong}}.} \bibinfo{year}{2006}\natexlab{}.
\newblock \showarticletitle{An Effective Approach for Periodic Web
  Personalization}. In \bibinfo{booktitle}{\emph{2006 IEEE/WIC/ACM
  International Conference on Web Intelligence (WI 2006 Main Conference
  Proceedings)(WI'06)}}. \bibinfo{pages}{284--292}.
\newblock
\urldef\tempurl%
\url{https://doi.org/10.1109/WI.2006.36}
\showDOI{\tempurl}


\end{thebibliography}

\appendix
\appendix
\section{Enable/Disable API Evaluation Set}
\label{appendix}
Test set for evaluation of skills with identical invocation phrases (Sec. \ref{identical_eval_skills}) -
\begin{enumerate}
    \item \{'Work excuses', 'Work Excuse Generator'\} with invocation phrase: `work excuses'
    \item \{'A pat on the Back', 'A pat on the Back'\} with invocation phrase: `a pat on the back'
    \item \{'Stock Market', 'Stock Market', 'UPRO Market Price'\} with invocation phrase: `stock market'
    \item \{'Joke Master', 'Joke Master', 'Joke Master',\} with invocation phrase: `joke master'
    \item \{'Sevenstax Coffee Maker', 'Coffee Maker', 'Coffee Maker', 'Coffee Maker'\} with invocation phrase: `coffee maker'
    \item \{'My Home', 'pi home', 'MY HOME',  'My home cake'\} with invocation phrase: `my home'
    \item \{'Good Night', 'Sounds: Good Night', 'Good Night', 'Good Night', 'Goodnight With Motivation Success Quotes'\} with invocation phrase: `good night'
    \item \{'Inspiring Quotes', 'All Time Inspiring Quotes', 'Inspiring Quotes', 'Inspiring Quotes', 'Inspiring Quotes'\} with invocation phrase: `inspiring quotes'
\end{enumerate}
\end{document}